\documentclass[11pt,a4paper]{article}
\usepackage{jheppub}
\usepackage{graphicx}
\usepackage{lipsum}
\usepackage{slashed}
\usepackage{amsmath}
\usepackage[normalem]{ulem}
\usepackage{enumitem}\usepackage{mathtools}
\usepackage{bm}
\usepackage[all]{hypcap}
\usepackage{tikz}
\usepackage{tikz-feynman}
\usepackage{bbm}
\usepackage{xcolor}
\usepackage{subcaption}
\usepackage[font=footnotesize,labelfont=bf]{caption}
\usepackage{mathabx}
\usepackage{pifont}

\allowdisplaybreaks

\def\C{\mathbb{C}}

\newcommand{\rep}[1]{\mathbf{#1}}

\usepackage{xspace}

\newcommand{\be}{\begin{equation}}
\newcommand{\ee}{\end{equation}}
\newcommand{\bea}{\begin{eqnarray}}
\newcommand{\eea}{\end{eqnarray}}
\newcommand{\beq}{\begin{equation}}
\newcommand{\eeq}{\end{equation}}

\newcommand{\cL}{{\cal L}}

\begin{document}

\title{ Phenomenology of a Deconstructed Electroweak Force }

\author[a,b]{Joe Davighi,} 
\emailAdd{joseph.davighi@cern.ch}

\author[c]{Alastair Gosnay,}
\emailAdd{a.gosnay.1@research.gla.ac.uk}

\author[c]{David J Miller,}
\emailAdd{david.j.miller@glasgow.ac.uk}

\author[c]{and Sophie Renner}
\emailAdd{sophie.renner@glasgow.ac.uk}

\affiliation[a]{Physik-Institut, Universit\"at Z\"urich, 8057 Z\"urich, Switzerland}
\affiliation[b]{Theoretical Physics Department, CERN, 1211 Geneva, Switzerland}
\affiliation[c]{School of Physics and Astronomy, University of Glasgow, Glasgow G12 8QQ, United Kingdom}

\abstract{
We study an effective theory of flavour in which the $SU(2)_L$ interaction is `flavour-deconstructed' near the TeV scale. 
This arises, for example, in UV models that unify all three generations of left-handed fermions via an $Sp(6)_L$ symmetry.
Flavour-universality of the electroweak force emerges accidentally (but naturally) from breaking the $\prod_{i=1}^3 SU(2)_{L,i}$ gauge group to its diagonal subgroup, delivering hierarchical fermion masses and left-handed mixing angles in the process.  
The heavy gauge bosons transform as two $SU(2)_L$ triplets that mediate new flavour non-universal forces. The lighter of these couples universally to the light generations, allowing consistency with flavour bounds even for a TeV scale mass. Constraints from flavour, high mass LHC searches, and electroweak precision are then highly complementary, excluding masses below 9 TeV. The heavier triplet must instead be hundreds of TeV to be consistent with meson mixing constraints. Because only the lighter triplet couples to the Higgs, we find radiative Higgs mass corrections of a few hundred GeV, meaning this model of flavour is arguably natural. The natural region will, however, be almost completely covered by the planned electroweak programme at FCC-ee.
On shorter timescales, significant parameter space will be explored by the High-Luminosity LHC measurements at high-$p_T$, and upcoming lepton flavour violation experiments, principally Mu3e.
}

\begin{flushright}
 CERN-TH-2023-246   
\end{flushright}

\maketitle

\section{Introduction}

The matrices of Yukawa couplings in the Standard Model (SM), that govern the interactions between the Higgs field $H$ and three generations of quarks and charged leptons, are highly non-generic. The eigenvalues of these matrices are extremely hierarchical, with {\em e.g.} $y_u/y_t \sim 10^{-5}$, and the mixing angles needed to diagonalise the quark Yukawa matrices are small and also hierarchical, with $1\gg |V_{us}| \gg |V_{cb}| \gg |V_{ub}|$. The origin of these hierarchies constitutes the {\em SM flavour puzzle}. While small Yukawa couplings are technically natural, being protected by chiral symmetries, this rich structure likely finds its explanation via new dynamics beyond the SM (BSM). A BSM solution to the flavour puzzle will typically feature new particles that interact differently with the different generations.

Symmetries, in this case the approximate global symmetries of the SM, provide important clues that should guide our attempts to solve the flavour puzzle. Each 3-by-3 complex Yukawa matrix $Y_u$, $Y_d$, and $Y_e$ is, to zeroth order, dominated by a single large entry responsible for setting the size of $y_t$, $y_b$, and $y_\tau$. Ignoring all other entries, such a matrix is symmetric under $U(2)_L \times U(2)_R$ rotations acting on the first and second (henceforth `light') generation left- and right-handed fermions. The unequal light generation masses, and the observed mixings between left-handed quarks, provide small breaking of these $U(2)_L \times U(2)_R$ global symmetries~\cite{Kagan:2009bn,Barbieri:2011ci,Isidori:2012ts,Fuentes-Martin:2019mun}, rendering them only approximate. 

If there is light BSM physics with sizeable couplings to the SM, where `light' indicates a scale of order TeV that could be probed at colliders, then the new physics couplings to the SM must also exhibit $U(2)$ flavour symmetries -- whether that BSM is invoked to explain the SM flavour puzzle or not. This flavour symmetry is required because of precise measurements in the flavour sector that agree well with the SM predictions, particularly concerning kaon mixing, which constrains the effective scale $\Lambda_{sd}$ appearing in certain 4-quark operators $\sim (\bar{s}d)^2/\Lambda_{sd}^2$ to be at least $10^{5-6}$ TeV. The need to reconcile these stringent flavour bounds with light BSM is often termed the {\em BSM flavour puzzle}. If we remain optimistic that BSM physics could be light, then it is tempting to hypothesize that the global $U(2)$ symmetries appearing in both the SM Yukawa sector and in the BSM sector have a common dynamical origin, for example emerging as accidental symmetries arising from a gauge symmetry that is intrinsically flavour non-universal (acting differently on the third generation), broken somewhere near the TeV scale. 

It is not just optimism that favours the BSM physics being light. If it were heavy, then the Higgs mass squared parameter would receive large radiative corrections, scaling quadratically with the heavy mass scale, that would render the Higgs mass fine-tuned -- irrespective of what dynamics reside at even higher scales.
These finite Higgs mass corrections will typically be generated at low loop order (1- or 2-loop) in BSM theories of flavour, since these typically couple to at least the Higgs or the top quark. This notion of `finite naturalness'~\cite{Farina:2013mla} is an important consideration guiding low-scale model building, that we take seriously in the present work.

There are many different symmetries one might gauge that would deliver $U(2)$ global symmetries as accidental. The most direct is to gauge anomaly-free combinations of the $U(2)$ flavour symmetries themselves. To give a few examples, low-scale models have been proposed for gauged $SU(2)^3$ with one factor for each type of quark field~\cite{DAgnolo:2012ulg}, for choices of $U(2)$ compatible with $SU(5)$ grand unification~\cite{Linster:2018avp}, and recently for gauged $SU(2)_{q_L+\ell_L}$ that acts only on left-handed fermions~\cite{Greljo:2023bix} (see also~\cite{Darme:2023nsy}). In all these options the gauged symmetry is {\em horizontal}, meaning it commutes with the SM gauge symmetry. The BSM forces responsible for explaining flavour are therefore totally decoupled from the SM interactions which are flavour-universal. The new force has a gauge coupling that is a free parameter which can be very small, decoupling its phenomenological effects even if the mass scale is low. The first example of such a horizontal gauge model of flavour, due to Froggatt and Nielsen~\cite{Froggatt:1978nt}, was not based on $SU(2)$ symmetries but on gauging a non-universal $U(1)_F$. Charges can be chosen to engineer realistic Yukawa structures, which are generated (typically by integrating out chains of extra fermions) upon breaking $U(1)_F$.

An alternative approach to generating accidental $U(2)$ flavour symmetries, which is the one we explore in this paper, is instead to `deconstruct'~\cite{Arkani-Hamed:2001nha} the SM gauge interactions according to flavour (see {\em e.g.}~\cite{Li:1981nk,Dvali:2000ha,Craig:2011yk,Panico:2016ull}). That is, we entertain a symmetry breaking pattern
\begin{equation} \label{eq:intro-SSB}
    G_{1+2}\times G_3 \xrightarrow{v_{23}\,\sim\, \text{TeV}}  G_{1+2+3}\, ,
\end{equation}
where $G$ denotes some part of the SM gauge symmetry. If the Higgs is charged under the $G_3$ factor only, the Yukawa couplings will inherit exact $U(2)$ flavour symmetries accidentally, which are then approximately realised in the symmetry-broken SM phase. This option, in comparison with the horizontal approach described above, has several appealing features:
\begin{itemize}[leftmargin=*]
    \item The ultraviolet (UV) embedding of the SM gauge interactions themselves is intrinsically non-universal; the flavour puzzle is not `factorised' from the known SM forces, as in the horizontal approach. The new gauge couplings cannot be arbitrarily small; each of the $g_{12}$ and $g_3$ gauge couplings must be at least as large as the SM gauge coupling onto which they match. This means the extra gauge bosons cannot be decoupled phenomenologically when their mass is low, unlike for a horizontal gauge extension.
    \item The choice of symmetry, by which we mean the groups involved and the representations in which the SM fields transform, is entirely dictated by the SM gauge structure; there are, as such, no {\em ad hoc} choices to make concerning {\em e.g.} the assignment of $U(1)_F$ charges. Moreover, anomaly cancellation is automatically inherited from the SM.
    \item It is easy to find semi-simple UV completions of a deconstructed gauge model, through which one can also explain the quantisation of hypercharge, and perhaps even identify a model with asymptotically free gauge couplings (see {\em e.g.}~\cite[\S 7]{Davighi:2023iks}). At the very least, one can replicate the known semi-simple embeddings of the flavour-universal SM, {\em i.e.} via $SU(5)$, $\mathrm{Spin}(10)$, or the Pati--Salam embedding~\cite{Pati:1974yy}, for each generation. This is the approach taken in the `Pati--Salam cubed' model of~\cite{Bordone:2017bld}, and recently in an $SU(5)^3$  model of `tri-unification'~\cite{FernandezNavarro:2023hrf}. Models like this could in turn be realised in a 5d setup, {\em e.g.}~\cite{Fuentes-Martin:2022xnb}.
    There are also other intrinsically flavoured options, sticking to 4d, as categorised in Ref.~\cite{Allanach:2021bfe}, in which the generations are further `re-unified' in the UV via some `gauge-flavour-unification' symmetry~\cite{Davighi:2022vpl}.
    For example, given $N_g$ generations one can embed $SU(2)_L$ into $Sp(2N_g)_L=Sp(6)_L$, as used in the electroweak flavour unification model of~\cite{Davighi:2022fer}.
    In contrast, most horizontal $G_{\mathrm{SM}} \times U(1)_F$ gauge theories, even those that are anomaly-free, have no semi-simple completion~\cite{Davighi:2022dyq}.
    \item At least when the symmetry group $G$ is semi-simple, the breaking pattern (\ref{eq:intro-SSB}), by which two copies of the same group $G$ are broken to their diagonal (which in this context means `flavour-universal') subgroup, is generic. It is independent of both the initial gauge coupling strengths, and of the representation of the scalar condensate (provided, of course, it is not a singlet under either copy of $G$). This follows from Goursat's lemma~\cite{Goursat1889,bauer2015generalized} concerning the subgroups of a direct product group, as was recently shown in the context of clockwork theories~\cite{Giudice:2016yja} in Ref.~\cite{Craig:2017cda}.
\end{itemize}
Thus, the $U(2)$ flavour symmetries which help us to reconcile both the SM and BSM flavour puzzles, together with the flavour-universality of SM gauge interactions, emerge accidentally but naturally in a model with deconstructed gauge symmetry.

Interest in this class of flavour models was recently revived thanks to the intriguing hints of BSM in $B$-meson decays, principally the evidence of tau {\em vs.} light-lepton flavour universality violation in charged current $b \to c \ell \bar{\nu}$ decays (according to the latest HFLAV fit the significance of the discrepancy in the $R_{D^{(\ast)}}$ observables is about $3.3\sigma$~\cite{HeavyFlavorAveragingGroup:2022wzx}). So-called `4-3-2-1 models'~\cite{Buttazzo:2017ixm,DiLuzio:2017vat,Bordone:2017bld,Greljo:2018tuh,DiLuzio:2018zxy,Fuentes-Martin:2020bnh,Fuentes-Martin:2022xnb,FernandezNavarro:2022gst,Davighi:2022bqf}, which feature an $SU(3)_{1+2}\times SU(4)_3$ deconstruction of colour along with quark-lepton unification in the third family, predict a flavoured $U_1$ leptoquark with mass a few TeV, that can explain this and other long-standing anomalies in $B$-meson decays (such as in $b\to s\mu\mu$ processes).
In this paper we step back from the $B$-anomalies (which motivate the $U_1$ leptoquark), and consider only flavour. Then it is {\em electroweak deconstruction} that is essential, while deconstructing colour is not, simply because the Higgs is colourless.\footnote{That said, if we seek a more detailed low-scale deconstructed flavour model in which the spurions generating $V_{cb}$ and $y_2/y_3$ are disentangled, and we further postulate that the gauge model has a semi-simple completion without adding further fermions, then the viable models {\em do} also feature a deconstructed colour group, and in particular an $SU(4)_3$ force, as shown in~\cite{Davighi:2023iks}.  }
The option of deconstructing hypercharge only was investigated in Refs.~\cite{FernandezNavarro:2023rhv,Davighi:2023evx} (see also~\cite{Allanach:2018lvl,Allanach:2019iiy,Davighi:2021oel}). 
Here we consider the other electroweak factor to be the origin of the SM flavour hierarchies, and deconstruct the $SU(2)_L$ interaction. 

We study in detail the symmetry breaking pattern 
    \begin{align}
        SU(2)_{L,1} \times SU(2)_{L,2} \times SU(2)_{L,3} &\xrightarrow{v_{12}\,\sim\,\mathcal{O}(100\text{~TeV})} SU(2)_{L,1+2}\times SU(2)_{L,3} \\
         &\xrightarrow{v_{23}\,\sim\,\mathcal{O}(\text{few~TeV})} SU(2)_{L,1+2+3}\, ,
    \end{align}
with the Higgs being a doublet of $SU(2)_{L,3}$.
We study this as an effective field theory (EFT) of flavour, not specifying the particular UV dynamics {\em above} the high scale $v_{12}$ that we presume generates the Yukawa structure; our purpose is rather to
elucidate the phenomenology of the gauge sector associated with (\ref{eq:intro-SSB}). 
One possible UV completion is via the $Sp(6)_L$~\cite{Davighi:2022fer} electroweak flavour unification group described above. The EFT we study captures the dominant low-energy phenomenology associated with the breaking of $Sp(6)_L$ in that model, which is an important motivation for the present paper.

The idea of a deconstructed $SU(2)_L$ is not new, but goes back to work of Ma and collaborators~\cite{Li:1981nk,Li:1992fi,Ma:1987ds,Ma:1988dn,Muller:1996dj,Malkawi:1996fs} in the 1980s -- predating the discovery of the complete third generation and the full CKM mixing pattern, which now forms a cornerstone of our motivation. The model was in part motivated by a then-anomalous measurement of the lifetime of the not-long-discovered tau lepton at SLAC, and predicted signatures in $B$-meson mixing and $m_W$. This selection of observables are indicative of some of the important phenomenology that we study in this paper.
While the gauge group considered there was the same, the setup of the model at low energies was different, with Higgs doublets coupled to each generation (compared to only one light Higgs coupled to $SU(2)_{L,3}$ in our case) and three sets of link field (rather than two). And needless to say, the motivations, the experimental and theoretical context, and the relevant phenomenology, are significantly different now.

The symmetry breaking (\ref{eq:intro-SSB}) gives six heavy gauge bosons: (i) a heavy $SU(2)_L$ triplet $W_{12}$ with mass $\sim v_{12}$, that mediates flavour violation in the 1-2 sector, and (ii) a lighter $SU(2)_L$ triplet $W_{23}$ at $\sim v_{23}$, which couples differently to the third generation but universally to the light generations.
As anticipated, the $W_{23}$ triplet gives unavoidable 1-loop corrections to the Higgs mass squared, that scale like $g_{SM}^2 v_{23}^2/(16\pi^2)$ where $g_{SM}$ is the $SU(2)_L$ SM gauge coupling. But thanks to the $U(2)$ protection of its couplings to the light generations, this triplet is phenomenologically viable close to the TeV scale. The electroweak scale can therefore be natural in this deconstructed $SU(2)_L$ framework, despite the proliferation of electroweak gauge bosons. The stability of the electroweak scale in a toy model with deconstructed $SU(2)_L$ symmetry was scrutinised in Ref.~\cite{Allwicher:2020esa}.

All these features motivate a comprehensive phenomenological study of the deconstructed $SU(2)_L$ gauge model, which we undertake in this paper. We elucidate the interplay of current experimental bounds in constraining the natural parameter space, finding excellent complementarity between flavour, electroweak precision observables, and high $p_T$ LHC searches in leptonic final states. This resonates with the model-independent analysis of~\cite{Allwicher:2023shc}, and also with the findings of ~\cite{Davighi:2023evx,FernandezNavarro:2023rhv} which explored the deconstructed hypercharge case. There are nevertheless important phenomenological differences with that scenario, due to the presence of charged currents, the left-handed chiral structure, and the preservation of custodial symmetry.
We examine flavour observables of interest, such as $B_s$-mixing, $B_s \to \mu\mu$, $B \to K^{(*)} \nu \bar{\nu}$, tau physics, and lepton flavour violating (LFV) processes.
Finally, we explore how the landscape of such a flavour model will evolve in the medium-term future, thanks to the huge leap forward brought by FCC-ee, but also due to significant shorter-term advances from the High-Luminosity LHC, Belle II, and Mu3e.

The structure of the paper is as follows.
In \S \ref{sec:model} we set out the model, including the symmetry breaking pattern and corresponding gauge boson spectrum. In \S \ref{sec:naturalness} we consider the Higgs mass stability. In \S \ref{sec:SMEFT} we match onto the Wilson coefficients of the SM effective field theory (SMEFT), and then in \S \S \ref{sec:W12} and \ref{sec:W23} we derive the phenomenological constraints on the heavy and light $SU(2)_L$ gauge boson triplets respectively. In \S \ref{sec:FCC} we discuss the prospects at future experiments, before concluding.

\section{The Model} \label{sec:model}

\subsection{Flavour deconstruction for flavour hierarchies} \label{sec:deconstruction}

We study a simplified, effective model of flavour based on a deconstructed $\Pi_{i=1}^3 SU(2)_{L,i}$ 
%\jd{\sout{$SU(2)_L^3$} [it might be confused with the 3rd family $SU(2)$, rather than $SU(2)$ cubed]}
gauge symmetry, that is spontaneously broken to the flavour-universal $SU(2)_L$ of the Standard Model:
\begin{equation} \label{eq:SSB-pattern}
    SU(2)_{L,1} \times SU(2)_{L,2} \times SU(2)_{L,3} \to SU(2)_{L,\text{SM}}.
\end{equation}
The $i^\text{th}$-generation of left-handed SM fermions is charged in the doublet representation of $SU(2)_{L,i}$. We take the SM Higgs to be charged only under $SU(2)_{L,3}$. For simplicity, we leave $SU(3)$ colour and $U(1)_Y$ hypercharge flavour-universal, as they are in the SM.\footnote{Going deeper into the UV, one might wish to unify quarks and leptons via an $SU(4)$ colour group \`a la Pati and Salam, either flavour universally or non-universally, and/or flavour deconstruct hypercharge also. One UV setup that combines these elements utilises the gauge group $SU(4) \times Sp(6)_L \times Sp(6)_R$, as in Ref.~\cite{Davighi:2022fer}. We comment on this option in \S \ref{sec:EWFU}, but our primary focus is on the phenomenology of deconstructed $SU(2)_L$.
} 
The symmetry breaking (\ref{eq:SSB-pattern}) to the SM occurs due to the condensing of two scalar bi-fundamental link fields $\phi^{12}\sim (\rep{2},\rep{2}, \mathbf{1})$ and $\phi^{23}\sim (\mathbf{1} ,\mathbf{2},\mathbf{2})$ which take the vevs:
\begin{equation}
\label{eq:scalarvevs}
    \langle \phi^{12}_{a_1 a_2} \rangle =v_{12}\,\epsilon_{a_1 a_2}, ~~~~\langle \phi^{23}_{a_2 a_3} \rangle =v_{23}\,\epsilon_{a_2 a_3},
\end{equation}
where $a_i$ is an index labelling $\C^2$ vectors acted on by the fundamental representation of $SU(2)_{L,i}$. The field content of the model is summarised in Table~\ref{tab:fields}. We derive the spectrum of associated heavy gauge bosons in \S \ref{sec:GB-couplings}. Here, we begin by describing the Yukawa sector in such a deconstructed $SU(2)_L$ model.

Because the Higgs field is charged under the third family part of the gauge group, renormalisable Yukawa couplings are permitted by the gauge symmetry only for the third family left-handed fields. Using a convention in which $H$ has hypercharge $-1/2$, we have
\begin{equation}
    -\cL \supset 
    y_t^i\, \overline{q}_{L,3} H u_{R,i} + 
    y_b^i\, \overline{q}_{L,3} H^c d_{R,i} +
    y_\tau^i\, \overline{\ell}_{L,3} H^c e_{R,i}\, ,
\end{equation}
where $H^c = i\sigma_2 H^\ast$, and where each of $y^i_{t,b,\tau}$ is a 3-component complex vector. With these couplings alone, the 3-by-3 complex Yukawa matrices $Y_{u,d,e}^{ij}$, defined such that $\cL \supset Y_{u}^{ij} \overline{q}_{L,i} H u_{R,j}$ {\em etc}, have non-zero entries only in the third row. Each of these Yukawa matrices is of course rank-1, with the non-zero eigenvalues being the components $y^3_{t,b,\tau}$. The matrices are moreover diagonalised (so that only the $Y^{33}$ entries are non-vanishing) via unitary rotations only of the {\em right}-handed fields,
which remain unphysical as in the SM (because all forces acting on RH particles are flavour-universal and neutral-current).

The UV theory must contain additional heavy dynamics,\footnote{The new dynamics does not strictly have to be hierarchically heavy; another option is for it to contribute to the effective Yukawa operators only via loop-suppressed diagrams (as in {\em e.g.}~\cite{Greljo:2023bix}), with the extra suppression translating to a higher effective scale.} 
such as extra Higgs-like scalars or vector-like fermions, which has already been integrated out at higher scales $\Lambda_{12}$ and $\Lambda_{23}$ to generate the remaining Yukawa couplings (responsible for the first and second generation masses and the CKM mixing) via higher-dimensional operators. Without specifying this UV dynamics explicitly (though we review one option in \S \ref{sec:EWFU}), one can write down these higher-dimensional operators in the effective field theory (EFT) description:
\begin{align} \label{eq:Yukawa-EFT}
    -\cL \supset \,    \frac{\phi^{23} }{\Lambda_{23}}\, &\left(C_c^i\, \overline{q}_{L,2}H u_{R,i} + 
    C_s^i\, \overline{q}_{L,2}H^c d_{R,i} +
    C_\mu^i\, \overline{\ell}_{L,2} H^c e_{R,i}\right) \\
    + \,  \frac{\phi^{12}\phi^{23} }{\Lambda_{12}\Lambda_{23}}\,
    &\left(C_u^i\, \overline{q}_{L,1}H u_{R,i} + 
    C_d^i\, \overline{q}_{L,1}H^c d_{R,i} +
    C_e^i\, \overline{\ell}_{L,1} H^c e_{R,i}\right) \nonumber\, ,
\end{align}
where $C^i_{c,s,\mu,u,d,e}$ are six more complex 3-vectors, this time containing Wilson coefficients determined by the matching from the (unspecified) UV theory.
Defining the pair of small parameters
\begin{equation}
    \epsilon_{ij} := \frac{v_{ij}}{\Lambda_{ij}}, \qquad ij\in\{12,23\}\, ,
\end{equation}
we expect the following textures for the effective Yukawa matrices after the symmetry breaking (\ref{eq:SSB-pattern}) has occurred:
\begin{equation} \label{eq:yukawa-hierarchies}
    Y_{u,d,e} \sim
    \begin{pmatrix}
        \epsilon_{12}\epsilon_{23} & \epsilon_{12}\epsilon_{23} & \epsilon_{12}\epsilon_{23} \\
        \epsilon_{23} & \epsilon_{23} & \epsilon_{23} \\
        1&1&1
    \end{pmatrix}\, ,
\end{equation}
up to factors of the order-1 Yukawa couplings and Wilson coefficients.

%%%%%%%%%%%%%%%%%%%%%%%%%%
\begin{table}
\begin{center}
\begin{tabular}{|c||c|ccc|}
\hline
Field(s)  & $SU(3) \times U(1)_Y$ & $SU(2)_{L,1}$ & $SU(2)_{L,2}$ & $SU(2)_{L,3}$ \\
\hline
$\psi_{L,1}$  & $\times$ & $\rep{2}$ & $\rep{1}$ & $\rep{1}$\\
$\psi_{L,2}$  & $\times$ & $\rep{1}$ & $\rep{2}$ & $\rep{1}$\\
$\psi_{L,3}$ & $\times$ & $\rep{1}$ & $\rep{1}$ & $\rep{2}$\\
$\psi_{R,i}$ & $\times$ & $\rep{1}$ & $\rep{1}$ & $\rep{1}$\\
\hline
$H$ &  $\times$ & $\rep{1}$ & $\rep{1}$ & $\rep{2}$ \\
\hline
$\phi^{12}$ &  $(\rep{1}, 0)$ & $\rep{2}$ & $\rep{2}$ & $\rep{1}$ \\
$\phi^{23}$ &  $(\rep{1}, 0)$ & $\rep{1}$ & $\rep{2}$ & $\rep{2}$ \\
\hline
\end{tabular}
\end{center}
\caption{Representations of SM fields under the deconstructed $SU(2)_L$ gauge symmetry. Here, $\psi_{L(R)}$ denotes a quark or lepton left- (right-) handed fermion multiplet. The Higgs is charged under the $SU(2)_{L,3}$ factor. In the second column, $\times$ denotes that the corresponding field is charged as in the SM under $SU(3)\times U(1)_Y$, which is true for all fields in this model. The BSM scalar fields $\phi^{ij}$ in the final two rows are required to break the deconstructed $SU(2)_L$ down to the SM, and in so doing generate the hierarchical structure for the SM Yukawa couplings.
} \label{tab:fields}
\end{table}
%%%%%%%%%%%%%%%%%%%%%%%%%%
    
Yukawa matrices with this structure would be diagonalised by order-1 rotations on the right-handed fields, {\em i.e.} with large mixing angles, while the left-handed rotations are hierarchial with the following structure
\begin{equation} \label{eq:Vude}
    V_{u,d,e} \sim \begin{pmatrix}
        1 & \epsilon_{12} & \epsilon_{12} \epsilon_{23} \\
        \cdot & 1 & \epsilon_{23} \\
        \cdot & \cdot & 1
    \end{pmatrix}\, ,
\end{equation}
up to order-1 coefficients that depend on the details of the UV dynamics that give rise to the EFT operators in (\ref{eq:Yukawa-EFT}). The CKM matrix, which governs the flavour-changing interactions of the SM $W^\pm$ bosons, is $V=V_u V_d^\dagger$ as usual, and so we expect
\begin{equation}
    \epsilon_{12} \sim \lambda, \qquad \epsilon_{23} \sim |V_{cb}| \sim \lambda^2\, ,
\end{equation}
where $\lambda \approx 0.2$ is the Cabibbo angle. One then expects that $|V_{ub}|\sim \epsilon_{12}\epsilon_{23}\sim \lambda^3$, in line with the measured value. In the phenomenological analysis that follows, it is useful to define two limiting cases, in which the CKM mixing comes entirely from either the up- or down-quark sector:
\begin{itemize}
    \item Up-alignment: $V_u = \mathbb{I}$, $V_d = V^\dagger$
    \item Down-alignment: $V_u = V$, $V_d = \mathbb{I}$.
\end{itemize}
These two benchmarks allow us to study the phenomenology in two extreme cases; for example, there is maximal down-type (up-type) meson mixing in the up-alignment (down-alignment) scenario -- see \S\S \ref{sec:kaon_mixing} and \ref{sec:Bs_mixing}. 

Lastly, the structure (\ref{eq:yukawa-hierarchies}) also implies the fermion mass eigenvalues follow a similar hierarchy to the CKM angles, with $y_1/y_2 \sim \epsilon_{12}$ and $y_2/y_3\sim \epsilon_{23}$. This predicted hierarchy offers a good starting point for explaining the observed mass and mixing hierarchies, although some of the Wilson coefficients have to be $\mathcal{O}(0.1)$ in order to fit the measured values (for example, to get light enough $m_{e,d,u}$, which are suppressed with respect to $m_{\mu,s,c}$ by more than just a Cabibbo factor, and to get $m_{b,\tau}$ which are significantly smaller than $m_t$). This is equally the case for the horizontal $SU(2)$ gauge model recently proposed in~\cite{Greljo:2023bix} which, by acting non-universally only on the left-handed fields, also predicts Yukawa textures like (\ref{eq:yukawa-hierarchies}); in both cases, the huge hierarchies of $\mathcal{O}(10^{-6})$ present in the SM Yukawa sector have been traded for acceptable factors of $\mathcal{O}(0.1)$.
If desired, further ingredients can be included to `break' this link between the mass and mixing hierarchies, for example deconstructing also the right-handed interactions; we refer the reader to {\em e.g.}~\cite{Bordone:2017bld,Davighi:2022fer,Fuentes-Martin:2022xnb,Davighi:2022bqf} for more complete flavour model-building efforts in this direction.

\subsection{Unification in the UV
} \label{sec:EWFU}

One possible UV origin for the symmetry breaking pattern (\ref{eq:SSB-pattern}) that we explore in this work, and which is an important motivation for our study, is an $Sp(6)_L$ gauge symmetry that unifies all three generations of left-handed doublets into one fundamental field:
\begin{equation}
    \psi_{L,1} \oplus \psi_{L,2} \oplus \psi_{L,3} \hookrightarrow \rep{6} \text{~~~of~~~} Sp(6)_L\, ,
\end{equation}
realising electroweak flavour unification~\cite{Davighi:2022fer} (see also~\cite{Kuo:1984md}). The high-scale symmetry breaking $Sp(6)_L \to \prod_{i=1}^3 SU(2)_{L,i}$ is triggered by the vev of a real scalar field $S_L$ in the $\rep{14}$-dimensional antisymmetric 2-index irrep of $Sp(6)_L$~\cite{Davighi:2022fer}, while the fields $\phi^{12}$ and $\phi^{23}$ (Table~\ref{tab:fields}) that trigger the lower-scale breaking fit inside a second real $\rep{14}$-plet $\Phi_L$. 
The physical Higgs field $H$ of the deconstructed $SU(2)_L$ model, as listed in Table~\ref{tab:fields}, is also embedded in the $\rep{6}$ of $Sp(6)_L$ alongside other flavoured copies that are presumed to be heavier.
It was shown in~\cite{Davighi:2022fer} that integrating out these heavy Higgs components at scales $\sim \Lambda_{12,23}$ also offers an explicit UV origin for the EFT operators in (\ref{eq:Yukawa-EFT}) that generate the hierarchial Yukawa structure.

In this paper we keep in mind the $Sp(6)_L$ UV scenario as a `benchmark' when exploring the low-energy phenomenology, which is determined by the low-scale symmetry breaking chain (\ref{eq:SSB-pattern}). 
The important thing from our low-energy point of view is that $Sp(6)_L$ unification gives a matching condition on the $SU(2)_{L,i}$ gauge couplings, which is that
\begin{equation} \label{eq:Sp6match}
 g_1 = g_2 = g_3 \qquad \text{[$Sp(6)_L$ matching condition]}
\end{equation}
at the matching scale.
Going to low energies (relevant to the experimental bounds we will compute), this matching condition is only slightly corrected by RG running. 
This of course depends on the precise mass scales at which each heavy scalar field is integrated out, but to get a handle on this effect we can compute the 1-loop $\beta$-functions in the EFT of Table~\ref{tab:fields}:
\begin{equation}
    \beta_i := \frac{\partial g_i}{\partial \ln \mu} = -\frac{g_i^3}{16\pi^2} \gamma_i, \qquad \gamma_1=\frac{35}{6}, \quad \gamma_2=\gamma_3=\frac{34}{6} < \gamma_1\, ,
\end{equation}
meaning that if we start at $\mu=m_Z$ and `run up', $g_1$ runs slightly faster than $g_2$ and $g_3$. More correctly, if we start from the matching condition (\ref{eq:Sp6match}) at the $Sp(6)_L$-breaking scale, which we might take to be $10^4$ TeV or so, then upon running to the low-energy model we will obtain $g_1 > g_{2,3}$ at $\mu\sim m_Z$. But the departure from equal couplings is, numerically, a very small effect despite the large logarithm coming from running over five orders of magnitude in scale; for instance, if we assume all the extra fields are integrating out at $\mu=10^4$ TeV and we run down to $\mu = m_Z$ assuming the field content of Table~\ref{tab:fields}, we find $g_1$ is bigger than $g_{2,3}$ by less than $1\%$. We are therefore happy to take (\ref{eq:Sp6match}) as our approximate condition for embedding the EFT inside the $Sp(6)_L$ model at any relevant phenomenological scale.

Interestingly, when we account for all the current experimental bounds, we will find in \S \ref{sec:W23} that this scenario for the gauge couplings allows for the lightest viable mass scale for the heavy gauge bosons.

\subsection{Gauge boson spectrum and couplings} \label{sec:GB-couplings}

We now derive the spectrum of heavy gauge bosons coming from the symmetry breaking pattern (\ref{eq:SSB-pattern}) and their couplings to SM fields, which determine the phenomenology we wish to study.

%Upon breaking $SU(2)_L^3\to SU(2)_L$, two of the $SU(2)_L$ triplet gauge bosons become massive, leaving one triplet of gauge bosons massless. 

\subsubsection*{Gauge boson masses}

The gauge boson masses come from the kinetic terms for the scalar bifundamental fields $\phi^{12}$ and $\phi^{23}$ that condense to break the $\prod_i SU(2)_{L,i}$ symmetry, via the vevs (\ref{eq:scalarvevs}). These kinetic terms are
\begin{equation}
\label{eq:Lscalarkin}
    \mathcal{L}_{\text{kin}}=\text{Tr}\left[(D_\mu \phi^{12})^\dagger D^\mu \phi^{12} \right]+\text{Tr}\left[(D_\mu \phi^{23})^\dagger D^\mu \phi^{23} \right],
\end{equation}
where the covariant derivatives are, making the distinct $SU(2)_{L,i}$ indices ($a_i$) explicit,
\begin{align}
    (D_\mu\phi^{12})_{a_1 a_2}&=\partial_\mu \phi^{12}_{a_1 a_2}-i g_1 (W^I_{1\mu} \tau^I)_{a_1 b_1} \phi^{12}_{b_1 a_2}-i g_2 (W^I_{2\mu} \tau^I)_{a_2 b_2} \phi^{12}_{a_1 b_2},\\
    (D_\mu\phi^{23})_{a_2 a_3}&=\partial_\mu \phi^{23}_{a_2 a_3}-i g_2(W^I_{2\mu} \tau^I)_{a_2 b_2} \phi^{23}_{b_2 a_3}-i g_3 (W^I_{3\mu} \tau^I)_{a_3 b_3} \phi^{23}_{a_2 b_3}.
\end{align}
Here $\tau^I=\sigma^I/2$ where $\sigma^I$ are the Pauli matrices, $g_1$, $g_2$ and $g_3$ are the respective gauge couplings for $SU(2)_{L,1}$, $SU(2)_{L,2}$ and $SU(2)_{L,3}$, and $W^I_{1\mu}$, $W^I_{2\mu}$ and $W^I_{3\mu}$ are the corresponding gauge fields in the unbroken $\prod_i SU(2)_{L,i}$-symmetric phase. 
Expanding Eq.~\eqref{eq:Lscalarkin} about the vevs~\eqref{eq:scalarvevs} yields the gauge boson mass terms
\begin{equation}
    [\mathcal{L}_{\text{kin}}]_{\phi\,\to\, \langle \phi \rangle} = \frac{v_{12}^2}{2}|g_1 \vec{W}_1-g_2 \vec{W_2}|^2+ \frac{v_{23}^2}{2}|g_2 \vec{W}_2-g_3 \vec{W_3}|^2,
\end{equation}
where we have now written each gauge boson triplet as a 3-vector in $SU(2)_{L,i}$ space, and where $|\vec{A}|^2$ denotes the usual Euclidean length squared of such a vector $\vec{A}$. 

Defining a small parameter $x\equiv (v_{23}/v_{12})^2 \ll 1$, the Lagrangian can be written as a 9-by-9 quadratic form (see also~\cite{Allwicher:2020esa}):
\be \label{eq:mass-matrix}
[\mathcal{L}_{\text{kin}}]_{\phi \to \langle \phi \rangle} = \frac{v_{12}^2}{2}
\begin{pmatrix}
\vec{W}_1 & \vec{W}_2 & \vec{W}_3 
\end{pmatrix}
\begin{pmatrix}
g_1^2 & -g_1 g_2 & 0 \\
-g_1 g_2 & g_2^2(1+x) & -g_2 g_3 x \\
0 & -g_2 g_3 x & g_3^2 x
\end{pmatrix}
\begin{pmatrix}
\vec{W}_1 \\ \vec{W}_2 \\ \vec{W}_3 
\end{pmatrix}.
\ee
As expected, this matrix has vanishing determinant and so there is a zero eigenvalue, corresponding to the gauge boson combination that remains massless thanks to $SU(2)_{L}$-universal remaining unbroken.
The eigenstates of the mass matrix (\ref{eq:mass-matrix}) are
\begin{align}
\vec{W}_{\text{SM}} &\,\propto\, \frac{1}{g_1} \vec{W}_1 + \frac{1}{g_2} \vec{W}_2 +\frac{1}{g_3} \vec{W}_3\, \qquad (\text{massless~eigenstate})\, , \label{eq:SMeigenstate}\\
\vec{W}_{12} &\,\propto\,  \frac{1}{g_2} \vec{W}_1 - \frac{1}{g_1} \vec{W}_2\, , \label{eq:12eigenstate}\\
\vec{W}_{23} &\,\propto\,  -g_1 g_2^2 \vec{W}_1 - g_1^2 g_2 \vec{W}_2+ (g_1^2 g_3+g_2^2 g_3) \vec{W}_2 \, .\label{eq:23eigenstate}
\end{align}
In terms of the UV gauge couplings and the vevs, the masses of the two heavy gauge boson triplets are
\begin{equation}
\label{eq:Wmasses}
m_{12} = v_{12}\sqrt{g_1^2+g_2^2} , ~~~~~~~ m_{23} =v_{23} \sqrt{\frac{g_1^2g_2^2+g_2^2g_3^2+g_1^2g_3^2}{g_1^2+g_2^2}}.
\end{equation}
Having derived the masses, we next derive the couplings of these gauge fields. 

Before doing so, we remark that there are also non-vanishing 3- and 4-point vertices coupling the heavy gauge triplets to the massless SM triplet, that come (from the UV perspective) from expanding out the gauge field kinetic terms in terms of the mass eigenstates. These interactions can be repackaged, from the low-energy perspective,  into the kinetic terms for the massive triplets, viewed as transforming in the adjoint representation of the unbroken flavour-universal $SU(2)_L$ gauge symmetry. Accordingly, when matching onto SMEFT in \S \ref{sec:SMEFT} the effects of these gauge-gauge interactions are automatically incorporated.

\subsubsection*{Gauge coupling matching condition and parametrisation}

As we mentioned in the Introduction, it is a group theoretic fact that a product $\prod_i G_i$ of multiple copies of the same simple group will, under generic conditions, always break down to its diagonal subgroup~\cite{Goursat1889,bauer2015generalized}. A precise version of this statement was made in the context of the clockwork mechanism~\cite{Giudice:2016yja} in Ref.~\cite{Craig:2017cda}, but here we see its relevance for models of flavour-deconstructed gauge symmetries: regardless of the values of the gauge couplings $g_i$, the unbroken gauge group will always be the diagonal subgroup.\footnote{We could have also played with scalar link fields $\phi$ in other representations of $SU(2)_{L,i}$, with the same effect. Choosing the link fields to transform as bidoublets under pairs of $SU(2)_{L,i}$ factors, as we do, is simply the minimal choice. } Equivalently, the massless gauge bosons, which we identify with the SM $SU(2)_L$ gauge bosons, necessarily couple flavour-universally, as they must. 

In the broken phase, the gauge coupling of the unbroken diagonal group $g_{\text{diag}}$, to which the massless gauge boson combination couples, is given in terms of the gauge couplings $g_i$ of the individual factors $G_i$ by~\cite{Li:1981nk,Craig:2017cda}
\begin{align} \label{eq:g-matching} 
\frac{1}{g_{\text{diag}}^2}= \sum_i \frac{1}{g_i^2}\, .
\end{align}
In our case, we therefore have that the SM electroweak coupling $g_{SM}$ is given in terms of $g_1$, $g_2$, $g_3$ by
\begin{align}
\label{eq:gsm}
g_{SM}= \left(\frac{1}{g_1^2}+\frac{1}{g_2^2}+ \frac{1}{g_3^2} \right)^{-1/2}
=\frac{g_1g_2g_3}{\sqrt{g_2^2g_3^2+g_1^2g_2^2+g_1^2g_3^2}} .
\end{align}
The fact that $g_{SM}$ is an observed quantity means that $g_1$, $g_2$ and $g_3$ are not independent, but are constrained by the matching condition (\ref{eq:g-matching}). In particular, Eq. (\ref{eq:g-matching}) immediately implies that each gauge coupling satisfies $g_i \gtrsim g_{SM}$, meaning that none of the UV groups $SU(2)_{L,i}$ can be `weakly coupled' to the SM fields.

\begin{figure}
    \centering
    \includegraphics[height=0.42\textwidth]{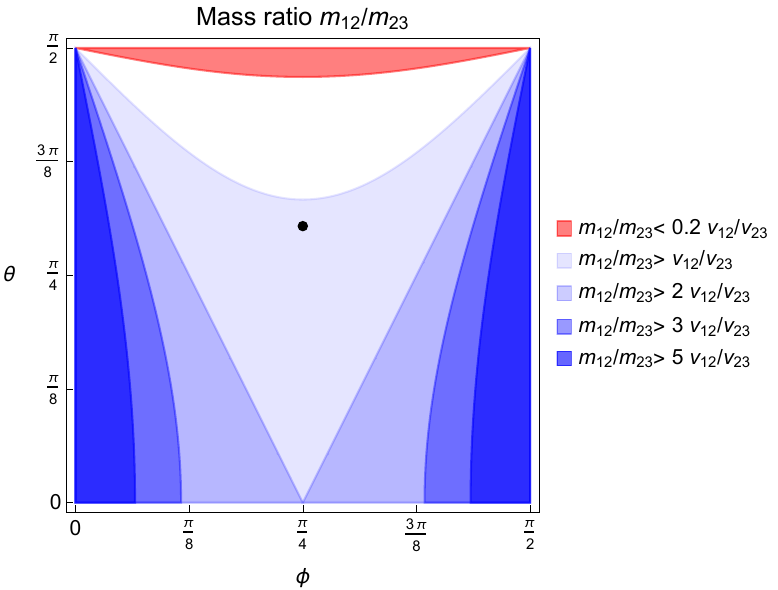}~~~~\includegraphics[height=0.42\textwidth]{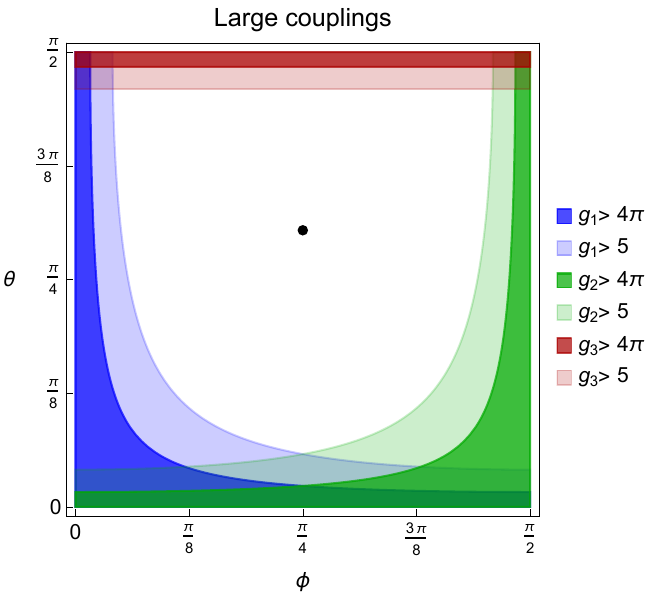}
    \caption{Variation of the masses and couplings of the heavy gauge bosons with the angles $\theta$ and $\phi$, as defined in Eq.~\eqref{eq:thetaphi}. \emph{Left:} regions corresponding to different ratios of the gauge boson masses $m_{12}/m_{23}$. \emph{Right:} regions with large values of the couplings $g_1$ (blue), $g_2$ (green) or $g_3$ (red). The black dots in both plots correspond to the points $g_1=g_2=g_3=\sqrt{3}g_{SM}$.}
    \label{fig:paramsvsangles}
\end{figure}

The constraint (\ref{eq:g-matching}) simply parametrizes a 2-sphere in 3d Cartesian coordinates $x_i = g_i^{-1}$, of radius $g_{SM}^{-1}$. It is therefore convenient to enforce this constraint by using polar coordinates $(\theta,\phi)$. We choose coordinates such that
\begin{equation}
\label{eq:thetaphi}
g_{SM}=g_3\cos\theta=g_2\sin\theta \cos\phi = g_1 \sin\theta \sin\phi\, ,
\end{equation}
which allow us to eliminate the couplings $g_{1,2,3}$ in terms of two physical mixing angles 
\begin{equation}
    \theta=\tan^{-1}\left(\frac{g_3}{g_1 g_2}\sqrt{g_1^2+g_2^2}\right)\, , \qquad 
    \phi=\tan^{-1}(g_2/g_1)\, .
\end{equation}
Without loss of generality, we restrict to the region $\theta, \phi \in [0,\frac{\pi}{2}]$ such that $g_1, g_2, g_3 > 0$. In these coordinates, the gauge boson masses in Eq.~\eqref{eq:Wmasses} become
\begin{equation}
\label{eq:Wmassespolar}
m_{12}=\frac{2 v_{12} g_{SM}}{\sin 2 \phi \,\sin \theta}, ~~~~~~m_{23}=\frac{2v_{23} g_{SM} }{\sin 2\theta}.
\end{equation}
%\textcolor{red}{Comment 1: are the physical values $\theta, \phi \in [0,\frac{\pi}{2}]$ - should there be an abs value here?} 
The hierarchy between the gauge boson masses therefore depends on the values of $\theta$ and $\phi$, as well as on the ratio between $v_{12}$ and $v_{23}$. This dependence is illustrated in the left plot in Fig.~\ref{fig:paramsvsangles}. On the right, regions of large couplings are shown on the same $\phi-\theta$ plane. By comparing the two plots, it is clear that the regions with the highest $m_{12}/m_{23}$ hierarchy are also the regions with large $g_1$ or $g_2$. This interplay is important when exploring the phenomenology of these gauge bosons, as we will see in later sections. The black dots in Fig.~\ref{fig:paramsvsangles} are the points at which $g_1=g_2=g_3=\sqrt{3}g_{SM}$, as expected in models where the $\Pi_{i=1}^3 SU(2)_{L,i}$ group itself arises from the breaking of an $Sp(6)_L$ group at a higher scale~\cite{Davighi:2022fer}, as outlined in \S \ref{sec:EWFU}.

\subsubsection*{Couplings to SM fermions}

By inverting and normalising Eqs.~\eqref{eq:SMeigenstate}-\eqref{eq:23eigenstate}, the UV `gauge eigenstate' fields $\vec{W}_1$, $\vec{W}_2$, and $\vec{W}_3$ can be written in terms of the mass eigenstates: 
\begin{align}
\vec{W}_1&=  \sin\phi \sin\theta\, \vec{W}_{SM} +\cos\phi\, \vec{W}_{12}+ \sin\phi \cos\theta\, \vec{W}_{23},\\
\vec{W}_2&= \cos\phi \sin\theta\, \vec{W}_{SM} -\sin\phi \, \vec{W}_{12}+ \cos\phi \cos\theta\, \vec{W}_{23},\\
\vec{W}_3&= \cos\theta \, \vec{W}_{SM} -\sin\theta \, \vec{W}_{23}.
\end{align}
These expressions allow us to find the couplings of the heavy gauge bosons to the SM fermions and Higgs.
First, we substitute into the fermionic Lagrangian to get the couplings to fermions. Note that, because we are flavour-deconstructing the $SU(2)_L$ interaction, we only have couplings to the left-handed SM fermions.
For the left-handed quark doublets we have
\begin{align}
    \mathcal{L}_{\text{quarks}}&=\sum_{i=1}^3 g_i \, (\bar{q}_L^{\prime i} {W}_{i}^{\mu I}\gamma_\mu \tau^I \, q_L^{\prime i}) \label{eq:lagdiffgs}\\
    &= g_{SM} {W}_{SM}^{\mu I}\sum_{i=1}^3 \, (\bar{q}_L^{\prime i} \gamma_\mu \tau^I \, q_L^{\prime i}) + \frac{g_{SM}}{\sin\theta}{W}_{12}^{\mu I} \left(\cot \phi \,(\bar{q}_L^{\prime 1} \gamma_\mu \tau^I \, q_L^{\prime 1}) -\tan \phi \,(\bar{q}_L^{\prime 2} \gamma_\mu \tau^I \, q_L^{\prime 2}) \right)\nonumber \\
    &+g_{SM} {W}_{23}^{\mu I}\left(\cot \theta \,(\bar{q}_L^{\prime 1} \gamma_\mu \tau^I \, q_L^{\prime 1}) + \cot \theta \,(\bar{q}_L^{\prime 2} \gamma_\mu \tau^I \, q_L^{\prime 2}) -\tan\theta\, (\bar{q}_L^{\prime 3} \gamma_\mu \tau^I \, q_L^{\prime 3}) \right)
\end{align}
where the prime on $q_L^\prime$ here recognises the fact that the gauge-flavour eigenstate fermion fields are in general not aligned with the fermion mass eigenstates (which will henceforth be denoted without primes, {\em e.g.} $q_L$); as usual the mass eigenstates are obtained by diagonalising the Yukawa interactions to the Higgs. We see that the SM part is flavour-universal in this basis. Flavour violation in the SM currents is induced only in the charged current interactions due to the rotation to the fermion mass basis. The leptonic Lagrangian is exactly analogous. 

We now rotate to the mass basis for the left-handed fermions, via
\begin{align}
q_L=V_u q_L^\prime\, , ~~~~~~~ l_L=V_l l_L^\prime\, ,
\end{align}
where $V_u$ and $V_l$ are 3-by-3 unitary matrices, and where we have taken the quark doublets to be defined as $q_L=(u_L, V d_L)^T$ where $u_L$ and $d_L$ are the mass eigenstate fields, and where $V$ is the CKM matrix.\footnote{We could equivalently make the choice $q_L=(V^\dagger u_L,d_L)^T$, and $q_L=V_d q_L^\prime$, since $V_u V_d^\dagger\equiv V$.} The overall quark Lagrangian can be written in matrix form as
\begin{align}
\mathcal{L}_{\text{quarks}}&= g_{SM} {W}_{SM}^{\mu I} (\bar{q}_L^{i} \gamma_\mu \tau^I \, q_L^{i}) + (g^q_{\mathcal{W}})_{[12]ij}{W}_{12}^{\mu I} (\bar{q}_L^{i} \gamma_\mu \tau^I \, q_L^{j})
\nonumber \\
     &+ 
(g^q_{\mathcal{W}})_{[23]ij}{W}_{23}^{\mu I} (\bar{q}_L^{i} \gamma_\mu \tau^I \, q_L^{j})\, ,
\end{align}
where summation over repeated indices is implied and the Hermitian coupling matrices $(g^q_{\mathcal{W}})_{[12]}$ and $(g^q_{\mathcal{W}})_{[23]}$ are 
\begin{equation}
\label{eq:gqW12}
(g^q_{\mathcal{W}})_{[12]}=\frac{g_{SM}}{s_\theta c_\phi s_\phi}V_u \begin{pmatrix}
c_\phi^2 & 0 & 0 \\
0 & -s_\phi^2 & 0 \\
0 & 0 & 0
\end{pmatrix} V_u^\dagger\, ,
\qquad
(g^q_{\mathcal{W}})_{[23]}=\frac{g_{SM}}{s_\theta c_\theta}V_u \begin{pmatrix}
c_\theta^2  & 0 & 0 \\
0 & c_\theta^2 & 0 \\
0 & 0 & -s_\theta^2
\end{pmatrix} V_u^\dagger\, ,
\end{equation}
where $s_x:=\sin x$ and $c_x:=\cos x$.
The leptonic couplings $(g^l_{\mathcal{W}})_{[12]}$ and $(g^l_{\mathcal{W}})_{[23]}$ are analogous to their quark counterparts, with the substitution $V_u\to V_l$:
\begin{equation}
\label{eq:glW12}
(g^l_{\mathcal{W}})_{[12]}=\frac{g_{SM}}{s_\theta c_\phi s_\phi}V_l \begin{pmatrix}
c_\phi^2 & 0 & 0 \\
0 & -s_\phi^2 & 0 \\
0 & 0 & 0
\end{pmatrix} V_l^\dagger\, ,
\qquad
(g^l_{\mathcal{W}})_{[23]}=\frac{g_{SM}}{s_\theta c_\theta}V_l\begin{pmatrix}
c_\theta^2  & 0 & 0 \\
0 & c_\theta^2 & 0 \\
0 & 0 & -s_\theta^2
\end{pmatrix} V_l^\dagger \, .
\end{equation}

Finally, we derive the couplings to the Higgs. The SM Higgs is charged only under $SU(2)_{L,3}$, as recorded in Table~\ref{tab:fields}, in order to explain the hierarchical heaviness of the third generation (\S \ref{sec:deconstruction}). Its covariant derivative therefore contains the pieces
\begin{align}
    D^\mu H &\supset \partial^\mu H-i g_3 {W}_{3}^{\mu I} \tau^I H\\
    &= \partial^\mu H - i g_{SM} {W}_{SM}^{\mu I}\tau^I H +i g_{SM}\tan\theta \, {W}_{23}^{\mu I}\tau^I H 
\end{align}
Including also the coupling to the hypercharge gauge boson, which is untouched in the model, we have
\begin{equation}
    D^\mu H = D_{SM}^\mu H +i g_{SM}\tan\theta \, {W}_{23}^{\mu I}\tau^I H.
\end{equation}
The $W^\mu_{23}$ couplings to the Higgs are then found by expanding out the Higgs kinetic term:
\begin{align}
\mathcal{L}_H&=| D^\mu H|^2 \supset - g_{SM }\tan \theta\, \vec{W}_{23}^{\mu I}H^\dagger \tau^I i D_{SM}^\mu H + \text{h.c.}.
\end{align}
For future convenience, we hence define a coupling
\begin{equation}
\label{eq:gphidef}
    (g^H_{\mathcal{W}})_{[23]}=- g_{SM }\tan \theta.
\end{equation}
The $W^\mu_{12}$ gauge triplet, on the other hand, does not couple to third generation fields, hence it has no interactions with the Higgs. 

Finally, we emphasize that none of the couplings of the $W_{23}$ triplet have any dependence on the mixing angle $\phi$, which measures flavour violation in the 1-2 sector. Thus, when exploring the phenomenology of the $W_{23}$ triplet (which is light and hence provides the dominant contributions to most low-energy phenomenology), we can parametrise all effects and observables by considering only the $(m_{23},\theta)$ two-dimensional parameter space.

\section{Flavour deconstruction {\em vs.}~naturalness} \label{sec:naturalness}

The presence of heavy new particles coupled to the Higgs in our multi-scale model of flavour follows inevitably from deconstructing the electroweak symmetry -- which is itself necessary in order to explain the Yukawa hierarchies in this framework. In this case, the particles in question are the $W_{23}$ triplet of gauge bosons, which have direct couplings to the Higgs. 

In this Section we consider the stability of the Higgs mass squared parameter in the presence of this heavy layer of new physics, by computing the finite Higgs mass corrections arising from loops of these particles.
Following the principle of `finite naturalness'~\cite{Farina:2013mla}, which stipulates that these calculable radiative contributions to $m_H^2$ should not be excessively fine-tuned against eachother, we place constraints on the model parameter space that favour the $W_{23}$ triplet being as light as is viable. We find numerically similar constraints to those that were found for flavour-deconstructing the hypercharge interaction in~\cite{Davighi:2023evx}.
We also consider the impact of the scalar link fields $\phi_{12,23}$ that condense to break the symmetry down to the SM, which, like any BSM scalar field that gets a vev, also give a tree-level shift of the Higgs mass squared.

\subsection{Tree-level Higgs mass}

Given any extension of the SM scalar sector, one can always write down renormalisable cross-quartic interactions between any pair of scalar fields; when the extra scalars get vevs, as is the case here, then these cross-quartics give tree-level contributions to the Higgs mass squared.
The tree-level scalar potential contains terms
\begin{equation} \label{eq:mH-tree}
    V \supset m_H^2 |H|^2 + \lambda_{23} |\phi_{23}|^2 |H|^2 + \lambda_{12} |\phi_{12}|^2 |H|^2\, .
\end{equation}
So, the `cross-quartic' interactions here give tree-level contributions to the Higgs mass-squared parameter (here written in the electroweak-unbroken phase) after the link fields $\phi_{12,23}$ get their vevs:
\begin{equation} \label{eq:tree-level-mH}
    [m_H^2]_{(0)} = m_H^2 + \lambda_{23} v_{23}^2 + \lambda_{12} v_{12}^2 \, .
\end{equation}
If the couplings $\lambda_{12,23}$ are order-1, then these tree-level contributions must be fine-tuned against eachother, to deliver a physical $[m_H^2]_{(0)} \approx -(100 \text{~GeV})^2$ in the low-energy theory. For the $\lambda_{23}$ coupling this would be a tuning at the percent level, since $v_{23} \sim \mathcal{O}(\text{TeV})$, which is the familiar `little hierarchy' tuning in the presence of TeV scale new physics coupled to the Higgs. But if $\lambda_{12}\sim 1$, this would imply a huge tuning in the tree-level Higgs mass squared parameter, because meson mixing requires $v_{12} \sim \mathcal{O}(100\text{~TeV})$, as we will see in \S \ref{sec:W12}. One might simply accept this tuning between the tree-level parameters of the theory; it is, at least, not signalling any particular instability in the model parameter space. More appealingly, however, this fine-tuning can be avoided if the cross-quartic couplings, in particular $\lambda_{12}$, are suitably small.

\begin{figure*}[t]
\begin{center}
\begin{tikzpicture}
\begin{feynman}
\vertex (a) {$\phi_{23}$} ;
\vertex [right=0.5in of a] (b);
\vertex [right=0.7in of b] (c);
\vertex [right=0.4in of c] (d) {$\phi_{23}$};
\vertex [below=0.6in of a] (e) {$H$};
\vertex [below=0.6in of b] (f);
\vertex [below=0.6in of c] (g);
\vertex [below=0.6in of d] (h) {$H$};
%\node at (b) [circle,fill,inner sep=1.5pt,label=below:{ $\,\,\,\, g_{SM}\tan\theta$}]{};
\node at (b) [circle,fill,inner sep=1.5pt]{};
\node at (c) [circle,fill,inner sep=1.5pt]{};
\node at (f) [circle,fill,inner sep=1.5pt]{};
\node at (g) [circle,fill,inner sep=1.5pt]{};
\diagram* {
(a) -- [scalar] (b) -- [scalar] (c) -- [scalar] (d),
(e) -- [scalar] (f) -- [scalar] (g) -- [scalar] (h),
(b) -- [photon, edge label = $W_{23}$] (f),
(c) -- [photon, edge label = $W_{23}$] (g),
};
\vertex [right=0.5in of h] (i);
\vertex [above=0.2in of i] (j) {$\implies$};
\vertex [right=0.5in of i] (k) {$H$};
\vertex[right=0.6in of k] (l);
\node at (l) [square dot,fill,inner sep=3.5pt]{};
\vertex[right=0.3in of l] (l');
\vertex[above=0.4in of l'] (L) {$\langle\phi_{23}\rangle$}; 
\vertex[left=0.6in of L] (L') {$\langle\phi_{23}\rangle$};
\vertex[right=0.5in of l] (m) {$H$};
\diagram* {
(k) -- [scalar] (l) [blob] -- [scalar] (m),
(l) -- [scalar] (L),
(l) -- [scalar] (L'),
};
\end{feynman}
\end{tikzpicture}
\end{center}
\caption{One-loop contribution to the cross-quartic coupling $\lambda_{23}$ between $H$ and $\phi_{23}$ (left), and the induced tree-level shift in the Higgs mass squared upon $\phi_{23}$ acquiring its symmetry-breaking vev (right). 
\label{fig:1loop-quartic} }    
\end{figure*}
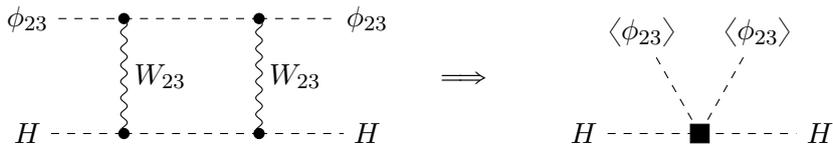

We should then ask if tuning $\lambda_{ij}$ close to zero is itself radiatively stable. Even if we set the coupling $\lambda_{23}$ to zero at some scale, it will be radiatively generated at 1-loop thanks to the $W_3^I$ gauge bosons, which couple to both $\phi_{23}$ and $H$ at tree-level, running in a box diagram (see Fig.~\ref{fig:1loop-quartic}). So, a radiatively stable estimate for the size of this coupling is\footnote{A similar argument was used to estimate quartic couplings in a scalar leptoquark model addressing the $B$-anomalies in Ref.~\cite{Davighi:2020qqa}.
}
\begin{equation}
    \lambda_{23} \sim \frac{g_{SM}^4 \tan^4\theta}{16 \pi^2} \ln{\frac{\mu^2}{m_{23}^2}} \sim 10^{-3}\, ,
\end{equation}
assuming the logarithm is not too large.
Under this assumption, the Higgs mass shift in (\ref{eq:tree-level-mH}) due to $v_{23}$ can even be in the $\mathcal{O}(100\text{~GeV})$ ballpark.

What about the contribution coming from the vev of the $\phi_{12}$ scalar?
While the contribution from $v_{12}$ is na\"ively more severe because $v_{12} \gg v_{23}$, the quartic coupling $\lambda_{12}$ is not generated at 1-loop order because the Higgs is neutral under $SU(2)_{L,1} \times SU(2)_{L,2}$, and so $H$ and $\phi_{12}$ do not talk to the same gauge bosons at tree-level. So, na\"ively accounting for at least one extra loop factor in suppression, we expect a value $\lambda_{12} \sim 10^{-5}$ to be radiatively stable. In such a scenario, {\em i.e.} if we assume the cross-quartics are mostly radiatively generated, the fine-tuning of the contributions (\ref{eq:tree-level-mH}) to $m_H^2$ need be no worse than the ten percent level. This kind of mechanism, whereby the Higgs is `shielded' from the higher scales in the theory associated with the 1-2 flavour violation (but without the need for {\em e.g.} SUSY or compositeness at low-scale), echoes the findings of Refs.~\cite{Allwicher:2020esa,Davighi:2023iks} that such a multi-scale setup can be radiatively stable.

\subsection{Radiative corrections to the Higgs mass}

Having dealt with the tree-level Higgs mass corrections coming from the scalar cross-quartics, we should then ask about additional loop contributions to the Higgs mass squared parameter, and thus complete our discussion of its radiative stability. There are 1-loop contributions to the Higgs 2-point function coming from both the scalar field $\phi_{23}$ and from the gauge bosons $W_{23}$ running in the loop. 

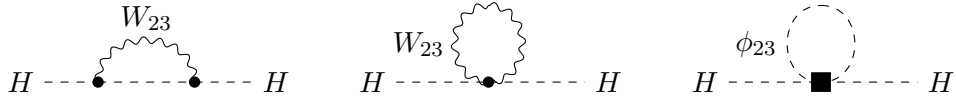
\begin{figure*}[t]
\begin{center}
\begin{tikzpicture}
\begin{feynman}
\vertex (a1) {$H$};
\vertex [right=0.4in of a1] (b);
\vertex [right=0.5in of b] (c);
\vertex [right=0.3in of c] (d) {$H$};
%\node at (b) [circle,fill,inner sep=1.5pt,label=below:{ $\,\,\,\, g_{SM}\tan\theta$}]{};
\node at (b) [circle,fill,inner sep=1.5pt]{};
\node at (c) [circle,fill,inner sep=1.5pt]{};
\diagram* {
(a1) -- [scalar] (b) -- [photon, half left, edge label = $W_{23}$] (c) -- [scalar] (d),
(b) -- [scalar] (c),
};
\vertex [right=0.5in of d] (A) {$H$};
\vertex [right=0.6in of A] (B);
\vertex [right=0.5in of B] (C) {$H$};
\vertex [above=0.4in of B] (D);
\node at (B) [circle,fill,inner sep=1.5pt]{};
\diagram* {
(A) -- [scalar] (B) -- [scalar] (C),
(B) -- [photon, half left, edge label = $W_{23}$] (D) -- [photon, half left] (B),
};
\vertex [right=0.5in of C] (A') {$H$};
\vertex [right=0.6in of A'] (B');
\vertex [right=0.5in of B'] (C') {$H$};
\vertex [above=0.4in of B'] (D');
\node at (B') [square dot,fill,inner sep=3.5pt]{};
\diagram* {
(A') -- [scalar] (B') -- [scalar] (C'),
(B') -- [scalar, half left, edge label = $\phi_{23}$] (D') -- [scalar, half left] (B'),
};
\end{feynman}
\end{tikzpicture}
\end{center}
\caption{One-loop contributions to the Higgs propagator generated by the gauge triplet $W_{23}$ (left two diagrams), and by the scalar $\phi_{23}$ given the cross-quartic coupling $\lambda_{23}$ ($\sqbullet$) which itself is radiatively generated as in Fig.~\ref{fig:1loop-quartic}. The heavier $W_{12}$ gauge bosons only give 2-loop Higgs mass corrections that are further suppressed by light fermion Yukawa couplings.
\label{fig:1loop-gauge} }    
\end{figure*}

\paragraph{1-loop scalar correction.}
The former contribution (right-most diagram of Fig.~\ref{fig:1loop-gauge}) takes the form
\begin{equation}
    [\delta m_H^2]_{(1)} \sim \frac{\lambda_{23}}{16 \pi^2} m_{\phi_{23}}^2\, .
\end{equation}
We expect this contribution to always be sub-leading compared to the tree-level cross-quartic contribution (\ref{eq:mH-tree}) we have already discussed: this is true as long as $m_{\phi_{23}}$ is not a whole factor $4\pi$ heavier than $v_{23}$, which would itself imply a non-perturbatively large quartic $|\phi_{23}|^4$ interaction. So, we can ignore this contribution, and focus instead on the gauge boson contributions to the Higgs propagator, which are parametrically different and are, importantly, more directly tied to the phenomenology of our flavour model.

\paragraph{1-loop gauge correction.}

First and foremost, there are 1-loop corrections to $m_H^2$ coming from the $W_{23}$ triplet of gauge bosons.
This 1-loop gauge contribution is the most important, being quadratically sensitive to the heavy mass scale (in this case $m_{23}$, the mass of the gauge triplet) determining the low-energy phenomenology of the model. It is also a numerically large contribution in the viable regions of parameter space. We therefore compute this contribution precisely (in contrast to our estimates above), to formulate a naturalness `constraint' on our model.

We find the finite contribution to the Higgs mass squared by computing the 1-loop contributions to the Higgs propagator with the heavy gauge triplet $W_{23}$ running in the loop (left two diagrams of Fig.~\ref{fig:1loop-gauge}), and evaluating at zero momentum. Using \texttt{Package-X}~\cite{Patel:2015tea} to do the loop integrals, we obtain
\begin{equation} \label{eq:mH-1loop}
     [\delta m_H^2]_{(1)} =   \frac{3 \tan^2 \theta  }{64\pi^2} g_{SM}^2\,  m_{23}^2\, \left(1+3\ln \frac{\mu^2}{m_{23}^2}\right)\, .
\end{equation}
Unlike the contributions from the scalar sector that we already discussed, which all depend on the otherwise undetermined cross-quartic couplings $\lambda_{ij}$, this contribution from the gauge sector scales like the gauge coupling which cannot be assumed small, but is rather matched onto the measured $SU(2)_L$ gauge coupling. So this gauge boson contribution to $m_H^2$ cannot be decoupled, and is moreover a calculable, finite function of our model parameters $(m_{23}, \theta)$.
If we take the RG scale $\mu = m_{23}$, cancelling the contribution from the log, then we can turn (\ref{eq:mH-1loop}) into a naturalness `constraint' on our parameter space by demanding:
\begin{align}
    |\tan\theta|\, m_{23} \lesssim \frac{8\pi}{\sqrt{3}\, g_{SM}} (\delta m_H)^{\text{max}} \approx 23 \, (\delta m_H)^{\text{max}}  \, ,
\end{align}
where $(\delta m_H)^{\text{max}}$ is the largest correction to $m_H$ that we deem tolerable.
% Depending on how aggressive we wish to be about the tolerable level of tuning, we can infer naturalness `constraints' on our parameter space:
% \begin{align}
%     |\tan\theta|\, m_{23} \lesssim \frac{8\pi}{\sqrt{3}\, g_{SM}} (\delta m_h)^{\text{max}} \approx 23 \, (\delta m_h)^{\text{max}}  \, .
% \end{align}
Allowing for a greater or lesser degree of tuning, we infer naturalness bounds:
\begin{align} \label{eq:natural}
    |\tan\theta|\, m_{23} \lesssim \begin{cases}
        2.8 \text{~TeV} \quad &\text{(no fine-tuning tuning)}\\
        23 \text{~TeV}
        &\text{(per-cent tuning in $m_H^2$)}\, .
    \end{cases}
\end{align}
For the `no tuning' benchmark we require $\delta m_H^2 \leq (125\text{~GeV})^2$, while for the looser bound we allow tuning $\delta m_H^2 \leq (\text{TeV})^2$, consistent with the usual `little hierarchy' between $m_H$ and the TeV. We will find (\S \ref{sec:combination}, Fig.~\ref{fig:W23_summary_current}) that the former `no tuning' benchmark is completely excluded by current experimental bounds on the $W_{23}$ triplet (consistent with the inference of a `little hierarchy', as is the case for most models), while there is plenty of viable parameter space consistent with the little hierarchy of $\mathcal{O}(\text{TeV})$ tuning in $|m_H|$.

\paragraph{2-loop gauge correction.}

We should also consider sensitivity of the Higgs mass to the even higher energy scale $m_{12}$, the mass of the $W_{12}$ triplet of gauge bosons. Because the Higgs is not charged under $W_{12}$, there is no 1-loop correction. The leading order correction is a 2-loop diagram, with a light-generation fermion running in the loop that emits a virtual $W_{12}$. Parametrically, this 2-loop diagram scales as
\begin{equation}
    [\delta m_H^2]_{(2)} \sim \left(\frac{1}{16\pi^2}\right)^2  \frac{g_{SM}^2 \tan^2\phi}{\sin^2\theta} \,\,y_2^2\, m_{12}^2\, \,,
\end{equation}
where $y_2$ is a second-generation Yukawa coupling, the largest of which is $y_c = 2 \times 10^{-3}$ (evaluated at an RG scale of order $10^3\text{~TeV}$~\cite{Greljo:2023bix}, of order $m_{12}$).
If we compare this correction to the 1-loop correction from $W_{23}$,
the extra loop factor plus the Yukawa suppression is more than enough to compensate the large ratio of masses $m_{12}/m_{23} \gtrsim 10^{1-2}$, and so we take the $W_{12}$ contributions to be subleading. Again, this is a manifestation of how the Higgs mass can be protected from the high scales associated with generating the 1-2 Yukawa sector~\cite{Allwicher:2020esa,Davighi:2023iks}, because the Higgs talks directly only to the third generation sector.

\paragraph{UV dependent corrections.}

Finally, in any UV complete model that matches onto our deconstructed $SU(2)_L$ EFT of flavour, there will of course be other contributions to the Higgs mass parameter 
coming from the UV dynamics that we assume generates the light Yukawa couplings. One possible origin for these couplings is to integrate out vector-like fermions; the consequences for naturalness due to such states, in similar-spirited deconstructed flavour models, have been discussed in Refs.~\cite{Davighi:2023evx,Davighi:2023iks}. An alternative extra ingredient might be extra Higgses, which can generate the light fermion masses via small mixing in the scalar sector -- additional loop corrections to the Higgs mass, and the stability of the scalar hierarchies in this scenario, were considered in {\em e.g.}~\cite{Allwicher:2020esa}.
Sticking with our EFT description, the naturalness estimate and constraint (\ref{eq:natural}) that we take here thus quantifies the `minimal tuning' coming unavoidably from the deconstructed gauge sector alone.

\section{SMEFT Matching} \label{sec:SMEFT}

To analyse the phenomenology of the heavy gauge bosons, it is convenient to first match their effects onto the coefficients of SMEFT operators. We use the Warsaw basis for dimension-6 SMEFT operators, introduced in Ref.~\cite{Grzadkowski:2010es}.
Using the tree-level dictionary in Ref.~\cite{deBlas:2017xtg}, we find tree level contributions to the Wilson coefficients of 4-fermion, Higgs-fermion, and 4-Higgs operators. Both heavy new gauge bosons, $W_{12}$ and $W_{23}$, match onto to 4-fermion operators, while only $W_{23}$ induces contributions to operators involving Higgs doublets. 

\paragraph{Four-fermion operators.}
The contributions from both gauge boson $SU(2)_L$ triplets to each of $C_{ll}$, $C_{qq}^{(3)}$, and $C_{lq}^{(3)}$ are, at tree level:
\begin{align}
(C_{ll})_{ijkl}&=\frac{1}{4m_{23}^2}\left( - (g^l_{\mathcal{W}})_{[23]kj}(g^l_{\mathcal{W}})_{[23]il}+\frac{1}{2}(g^l_{\mathcal{W}})_{[23]kl}^*(g^l_{\mathcal{W}})_{[23]ij}\right) + \{[23] \to [12] \}\, , \label{eq:Cll}\\
(C_{qq}^{(3)})_{ijkl}&=-\frac{(g^q_{\mathcal{W}})_{[23]kl}^*(g^q_{\mathcal{W}})_{[23]ij}}{8m_{23}^2}+ \{[23] \to [12] \}\, , \label{eq:Cqq}\\
(C_{lq}^{(3)})_{ijkl}&=-\frac{(g^q_{\mathcal{W}})_{[23]kl}^*(g^l_{\mathcal{W}})_{[23]ij}}{4m_{23}^2}+ \{[23] \to [12] \}\, .\label{eq:Clq}
\end{align}
The couplings $g_{\mathcal{W}}^{l,q}$ are given in eqs.~\eqref{eq:gqW12} and \eqref{eq:glW12}, and $\lbrace i,j,k,l \rbrace$ label flavour indices.

\paragraph{Higgs-fermion operators.}
Only the lighter $W_{23}$ triplet of gauge bosons matches to Higgs-fermion operators, since $W_{12}$ has no tree-level coupling to the Higgs boson. The matching results are:
\begin{align}
(C_{H l}^{(3)})_{ij}&=-\frac{(g^H_{\mathcal{W}})_{[23]}(g^l_{\mathcal{W}})_{[23]ij}}{4m_{23}^2}\, ,\label{eq:CHl3}\\
(C_{H q}^{(3)})_{ij}&=-\frac{(g^H_{\mathcal{W}})_{[23]}(g^q_{\mathcal{W}})_{[23]ij}}{4m_{23}^2}\, ,\label{eq:CHq3}
\end{align}
where the coupling $(g^H_{\mathcal{W}})_{[23]}$ is defined in Eq.~\eqref{eq:gphidef}.

\paragraph{Higgs operators.}
Lastly, we generate operators involving only Higgs doublets, again from integrating out the $W_{23}$ triplet. The non-zero Wilson coefficients generated at tree-level are:
\begin{align}
C_H &= - \frac{\lambda_{H}(g^H_{\mathcal{W}})_{[23]}^2}{m_{23}^2}\, ,\\
C_{H \Box} &= \frac{3(g^H_{\mathcal{W}})_{[23]}^2}{8m_{23}^2}\, ,
\end{align}
where $\lambda_{H}$ is the SM Higgs quartic coupling. Notice that the operator $C_{HD}$ is not generated at tree-level.

\section{Phenomenology of the high mass $W_{12}$ triplet} \label{sec:W12}

To reproduce the observed Yukawa structure we expect a hierarchical ratio of vevs $v_{12}/v_{23}\gg 1$, meaning that in general we expect the $W_{12}$ triplet to be significantly heavier than the $W_{23}$ triplet (see Eq.~\eqref{eq:Wmasses} and Fig.~\ref{fig:paramsvsangles}). Nevertheless, the phenomenology of the $W_{12}$ triplet can still lead to meaningful constraints on the model coming from {\em flavour}, because it can have significant flavour non-universal and/or flavour changing interactions within the first two fermion generations, which probe much higher scales than those directly explored by {\em e.g.} direct searches at the LHC. In this section we explore the implications of these.

\subsection{Kaon and $D$ meson mixing} \label{sec:kaon_mixing}
The neutral component of the $W_{12}$ triplet will induce unavoidable effects in at least one of neutral $K$- and $D$-meson mixing. 
The effective Lagrangian below the electroweak scale relevant for these mixing processes is:
\begin{equation}
\mathcal{L}_{\mathrm{eff}}\supset -C_1^K \left(\bar{d}_L \gamma^\mu s_L \right)^2 - C_1^D \left(\bar{u}_L \gamma^\mu c_L \right)^2.
\end{equation}
In general, $W_{12}$ will induce contributions to the Wilson coefficients (ignoring RG effects):
\begin{align}
C_{1}^K&=\frac{1}{8m_{12}^2} \frac{g_{SM}^2}{s_\theta^2 c_\phi^2 s_\phi^2} \left([V_d]_{11} [V_d]_{12}^*c_\phi^2-[V_d]_{21} [V_d]_{22}^*s_\phi^2\right)^2,\\
C_{1}^D&=\frac{1}{8m_{12}^2} \frac{g_{SM}^2}{s_\theta^2 c_\phi^2 s_\phi^2} \left([V_u]_{11}^* [V_u]_{21}c_\phi^2-[V_u]_{12}^* [V_u]_{22} s_\phi^2\right)^2.
\end{align}
%The rotation matrices must multiply to the CKM matrix: $V_uV_d^\dagger=V$. 
The relative strength of the $K$- {\em vs.} $D$-mixing bounds depends strongly on our particular alignment assumption.
We explore the limiting cases of \emph{up-alignment} ($V_u=\mathbb{I}$, $V_d=V^\dagger$) and \emph{down-alignment} ($V_d=\mathbb{I}$, $V_u=V$), as introduced in \S \ref{sec:deconstruction}. A realistic scenario is expected to lie between these two limits. 

\paragraph{Up-aligned scenario.}
In this case, $C_1^D=0$, while 
\begin{equation}C_{1}^{K,\text{up}}=\frac{1}{8m_{12}^2} \frac{g_{SM}^2}{s_\theta^2 c_\phi^2 s_\phi^2} \left(V_{ud}^* V_{us}c_\phi^2-V_{cd}^* V_{cs}s_\phi^2\right)^2\, .
\end{equation}
In the Wolfenstein parameterisation~\cite{Wolfenstein:1983yz} of the CKM matrix this becomes
\begin{align}
C_{1}^{K,\text{up}}&=\frac{1}{8m_{12}^2} \frac{g_{SM}^2}{s_\theta^2 c_\phi^2 s_\phi^2} \left(c_\phi^2+s_\phi^2\right)^2\lambda^2(1-\lambda^2)+ O(\lambda^6),\\
&=\frac{1}{8m_{12}^2} \frac{g_{SM}^2}{s_\theta^2 c_\phi^2 s_\phi^2}\lambda^2(1-\lambda^2)+O(\lambda^6),
\label{eq:upalignedKmix}
\end{align}
where $\lambda\approx 0.23$ is the Cabibbo angle. 

\paragraph{Down-aligned scenario.}
In this case, $C_1^K=0$, but
\begin{equation}
C_{1}^{D,\text{down}}=\frac{1}{8m_{12}^2} \frac{g_{SM}^2}{s_\theta^2 c_\phi^2 s_\phi^2} \left(V_{ud}^* V_{cd}c_\phi^2-V_{us}^*V_{cs}s_\phi^2\right)^2\, .
\end{equation}
Again using the Wolfenstein parameterisation of the CKM matrix, we here obtain
\begin{equation}
\label{eq:downalignedDmix}
C_{1}^{D,\text{down}}=\frac{1}{8m_{12}^2} \frac{g_{SM}^2}{s_\theta^2 c_\phi^2 s_\phi^2}\lambda^2(1-\lambda^2)+ O(\lambda^6)\, ,
\end{equation}
{\em i.e.} the same as $C_1^{K,\text{up}}$. Notice that both the contributions to $K$- and $D$-meson mixing observables are minimised when $\phi$ is $\pi/4$, corresponding to parameter space points where the gauge couplings $g_1$ and $g_2$ are equal. 

\paragraph{Experimental constraints.}
The real and imaginary parts of $C_1^K$ and $C_1^D$ are bounded by measurements of $K$- and $D$-meson mixing. In both the up- \eqref{eq:upalignedKmix} and down-aligned \eqref{eq:downalignedDmix} cases, it is clear by inspection that the induced Wilson coefficients are entirely real. For general $V_d$ (or $V_u$) there can be additional phases, but these will only induce observable imaginary parts of Wilson coefficients as factors in invariants suppressed by small Yukawas and CKM elements~\cite{Bonnefoy:2021tbt,Bonnefoy:2023bzx}, and are unlikely to lead to important constraints on the model.

\begin{figure}
    \centering
    \includegraphics[width=0.9\textwidth]{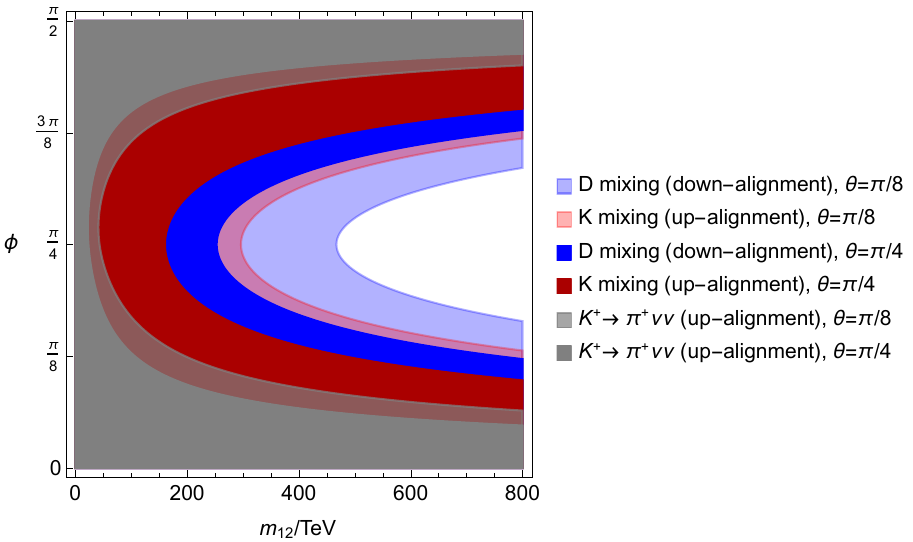}
    \caption{Bounds (at 95\% C.L.) on $m_{12}$ and $\phi$ from $K$- and $D$-meson mixing (\S \ref{sec:kaon_mixing}), and from the branching ratio of the rare kaon decay $K^+\to \pi^+ \bar{\nu} \nu$ (\S \ref{sec:Kpinunu}). In the \emph{up-alignment} (\emph{down-alignment}) scenario, the gauge eigenbasis of the $W_{12}$ couplings are aligned with the up (down) quark mass eigenstates.}
    \label{fig:KandDM12}
\end{figure}

The real parts of the Wilson coefficients are bounded as~\cite{SilvestriniLaThuile}
\begin{align}
\mathrm{Re} \, C_1^K &\in [-6.8,7.7] \times 10^{-13} \,\mathrm{GeV^{-2}},\\\mathrm{Re} \, C_1^D &\in [-2.5,3.1] \times 10^{-13} \,\mathrm{GeV^{-2}},
\end{align}
at 95\% C.L. These lead to constraints as shown in Fig.~\ref{fig:KandDM12}, for both up- and down-alignment, which we plot for two benchmark values of $\theta$, namely $\theta=\pi/8$ and $\theta=\pi/4$. While it can be seen from Eqs.~\eqref{eq:upalignedKmix} and \eqref{eq:downalignedDmix} that the meson mixing contributions are minimised for $s^2_\theta=1$ and hence $\theta=\pi/2$, since this is where $g_1$ and $g_2$ are minimised (see Eq.~\eqref{eq:thetaphi}), at this point $g_3$ diverges so it is not a physically relevant regions of parameter space. Instead, $\theta=\pi/4$ is an optimal benchmark point for the phenomenology of the $W_{23}$ gauge bosons, as we will see in the next section.

We infer from Fig.~\ref{fig:KandDM12} that the mass of the $W_{12}$ triplet must satisfy
\begin{equation} \label{eq:m12_bound}
    m_{12} \gtrsim 160 \text{~TeV}\, ,
\end{equation}
at least for the benchmark $\theta=\pi/4$ (for the $\theta=\pi/8$ benchmark the bounds are strictly stronger),
which we read off from the kaon mixing bound in the up-aligned scenario. Note that this bound is saturated at $\phi=\pi/4$, corresponding to the case where $g_1=g_2$, and we will see that this kaon mixing bound provides the strongest constraint on the $W_{12}$ parameter space. So, we learn that the bounds on $W_{12}$ are weakest for equal UV gauge couplings $g_1$ and $g_2$, as predicted for example by the $Sp(6)_L$ completion (\S \ref{sec:EWFU}).

\subsection{$K^+\to \pi^+\bar{\nu}\nu$} \label{sec:Kpinunu}
If $W_{12}$ has off-diagonal couplings to down-type quarks, it can also mediate $K^+\to \pi^+\bar \nu\nu$ decays at tree level. The effective Lagrangian to which both the SM and the $W_{12}$ matches is 
\begin{equation}
    \mathcal{L}=\frac{4 G_F}{\sqrt{2}}\frac{e^2}{16\pi^2 s_W^2}V^*_{td}V_{ts}C_L^i \left(\bar d \gamma^\mu P_L s \right)\left(\bar \nu_i \gamma_\mu (1-\gamma_5) \nu_i \right).
\end{equation}
From~\cite{Allwicher:2023shc,Buras:2015qea}, the branching ratio is
\begin{equation}
    \frac{BR(K^+\to \pi^+\bar{\nu}\nu)}{BR(K^+\to \pi^+\bar{\nu}\nu)_{SM}} = \frac{1}{3|\mathcal{A}_{SM}|^2}\sum_{e,\mu,\tau}\left|\Delta C_L^i +  \mathcal{A}_{SM}\right|^2\, ,
\end{equation}
where $\Delta C_L^i$ is the new physics contribution to the $C_L^i$ coefficient, and 
\begin{equation}
    \mathcal{A}_{SM}=C_{L,SM}^i + \frac{|V_{us}|^5|V_{cs}}{V^*_{td}V_{ts}}P_c
\end{equation}
is the full SM amplitude for the decay. This includes the short-distance SM piece $C_{L,SM}=-1.48\pm 0.01$~\cite{Buchalla:1998ba} and the long-distance part $P_c=0.404\pm 0.024$. 

The SM prediction is $BR(K^+\to \pi^+\bar{\nu}\nu)_{SM}=(9.11\pm 0.72)\times 10^{-11}$~\cite{Buras:2015qea}, while the experimental measurement from NA62 is $BR(K^+\to \pi^+\bar{\nu}\nu)_{exp}=(10.6\pm 3.8)\times 10^{-11}$~\cite{NA62:2021zjw}. In terms of the model parameters, we have 
\begin{equation}
\label{eq:KpinunuCLgen}
    \Delta C_L^i=\frac{4\pi v^2 s_W^2}{e^2 V^*_{td}V_{ts}} \frac{g_{SM}^2}{s_\theta^2 c_\phi^2 s_\phi^2} \left([V_d]_{11} [V_d]_{12}^* c_\phi^2-[V_d]_{21}[V_d]_{22}^*s_\phi^2 \right) \times \begin{cases}
   c_\phi^2 & (i=e),\\
    -s_\phi^2 & (i=\mu), \\
     0 & (i=\tau).
    \end{cases}
\end{equation}
If we are in the down-aligned scenario (such that $V_d=\mathbb{I}$), $\Delta C_L^i=0$, and we have no effects in $K^+\to \pi^+\bar \nu \nu$. In the up-aligned scenario ($V_d=V^\dagger$), Eq.~\eqref{eq:KpinunuCLgen} becomes
\begin{align}
   \Delta C_L^i=\frac{4\pi v^2 s_W^2}{e^2 V^*_{td}V_{ts}} \frac{g_{SM}^2}{s_\theta^2 c_\phi^2 s_\phi^2} \lambda \left(1-\frac{1}{2}\lambda^2 \right) \times \begin{cases}
   c_\phi^2 & (i=e),\\
    -s_\phi^2 & (i=\mu), \\
     0 & (i=\tau),
    \end{cases}
\end{align}
where $\lambda$ is the Wolfenstein parameter, and this equation is valid up to $O(\lambda^4)$. 

The resulting constraints on the parameter space of the $W_{12}$ in the up-aligned scenario are shown in grey in Fig.~\ref{fig:KandDM12}, where it can be seen that these are subdominant to the $K$ mixing constraints under the same assumptions. However, the SM prediction for the $K^+ \to \pi^+ \bar \nu \nu$ branching ratio is very theoretically clean, whereas the prediction of the real parts of the $K$ and $D$ mixing include long-distance contributions whose errors are difficult to quantify. The proposed HIKE experiment~\cite{HIKE:2023ext} at CERN stands to measure the $K^+ \to \pi^+ \bar \nu \nu$ branching ratio with a factor $\times 3$ improvement in precision over NA62, to obtain $5\%$ precision (after its first operation phase). However, up to the caveats already mentioned concerning SM theory uncertainties due to long-distance effects, we find that even with the HIKE improvements the constraints from meson mixing remain the leading constraint on the $W_{12}$ parameter space, giving the constraint  in Eq. (\ref{eq:m12_bound}).

\subsection{$\mu \to 3 e$} \label{sec:12_LFV}
Charged lepton flavour violation is not obligatory, since the rotation matrix $V_l$ in Eq.~\eqref{eq:glW12} is arbitrary, so the $W_{12}$ interactions could be exactly aligned to the mass eigenstates of the charged leptons. However, it is informative to understand how far we can stray from this alignment limit before encountering strong constraints.

We can parameterise the $V_l$ matrix as a rotation in three angles $\alpha$, $\beta$ and $\gamma$:
\begin{align} \label{eq:VlrotLargeAngles}
V_l&=\begin{pmatrix}
\cos \alpha & - \sin \alpha & 0\\
\sin \alpha & \cos \alpha & 0 \\
0 & 0 & 1
\end{pmatrix}
\begin{pmatrix}
\cos \beta & 0 & \sin \beta\\
0 & 1 & 0 \\
-\sin \beta & 0 & \cos \beta
\end{pmatrix}
\begin{pmatrix}
1 & 0 & 0 \\
0 & \cos \gamma  & -\sin \gamma\\
0 & \sin \gamma  & \cos \gamma
\end{pmatrix},\\
&\approx \begin{pmatrix}
1-\alpha^2/2 - \beta^2/2 & \beta \gamma - \alpha & \beta + \alpha \gamma \\
\alpha & 1-\alpha^2/2 - \gamma^2/2 & \alpha \beta - \gamma \\
- \beta & \gamma & 1-\beta^2/2 - \gamma^2/2
\end{pmatrix},
\label{eq:Vlrotsmallangles}
\end{align}
where the second line is an approximation up to second order in small angles. %\textcolor{red}{(Comment 4) The relative sign between $\gamma$ rotations and $\alpha$ and $\gamma$ rotations in (\ref{eq:VlrotLargeAngles}) is present since a 3$d$ Cartesian coordinate system is right-handed.}
In general, the matrix is unitary rather than orthogonal, but since the branching ratios that constrain lepton flavour violation are insensitive to phases, we approximate it as real. 

If the angle $\alpha\neq 0$, the $W_{12}$ will mediate $\mu \to 3 e$ decays at tree level through the four-lepton interaction Eq.\eqref{eq:Cll}. Using formulae from~\cite{Crivellin:2013hpa}, and the SMEFT coefficient~\eqref{eq:Cll}, this branching ratio is (up to second order in the small mixing angles $\alpha$, $\beta$, $\gamma$):
\begin{equation}
    BR (\mu \to 3 e) = \frac{m_\mu^5}{49152 \pi^3 \Gamma_\mu } \frac{g_{SM}^4 \alpha^2}{\sin^4\theta \sin^4\phi}.
\end{equation}
If we generically expect the $V_l$ matrix to be CKM-like, by analogy to the $V_{u,d}$ matrices, then we would expect the angle $\alpha$ to be similar in size to the Cabibbo angle, i.e.~$\sin\alpha \sim 0.2$.

\begin{figure}
    \centering
    \includegraphics[width=0.7\textwidth]{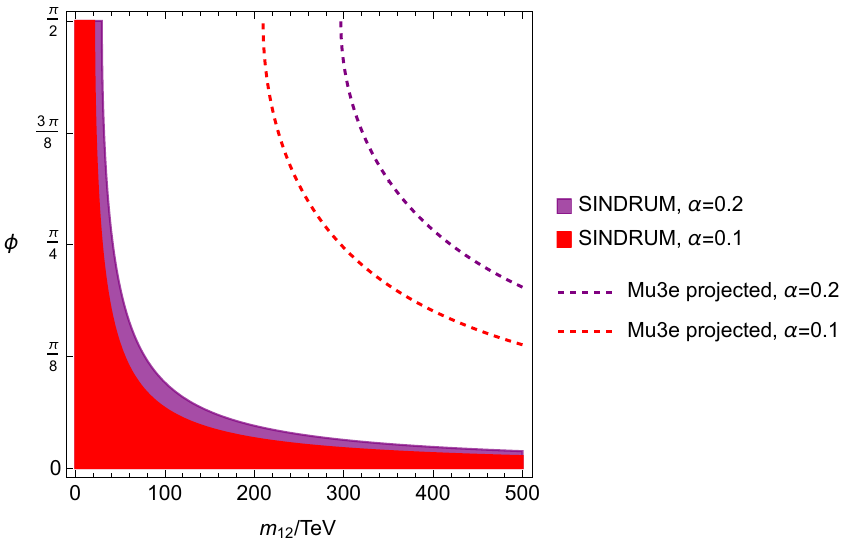}
    \caption{Regions of $W_{12}$ parameter space excluded (at 95\% C.L.) by the branching ratio of $\mu \to 3 e$, for two different values of the $\mu -e$ mixing angle $\alpha$. All these regions were calculated taking $\theta=\pi/4$. Dotted lines show projected limits from the future Mu3e experiment.}
    \label{fig:w12LFVphi}
\end{figure}

Currently, the best limit on the branching ratio of $\mu \to 3 e$ is $BR(\mu \to 3 e)<10^{-12}$ at 90\% C.L~from the SINDRUM experiment~\cite{SINDRUM:1987nra}. This limit is shown in Fig.~\ref{fig:w12LFVphi}, for two different values of $\alpha$, and taking $\theta= \pi/4$.\footnote{This value of $\theta$ has been chosen to be favourable for the phenomenology of the $W_{23}$, as we will see in the next Section.} It can be seen that with a CKM-like value of $\alpha=0.2$, current limits on $\mu \to 3 e$ do not impose strong constraints on the model when compared with the constraints from quark flavour found earlier in this section. On this plot we also show projected limits from the future Mu3e experiment, which are projected to reach a branching ratio of $10^{-16}$~\cite{Hesketh:2022wgw}. This will increase the limits on $m_{12}$ by one order of magnitude. Nevertheless, these projected constraints will still allow a CKM-like mixing between the first two generations in the lepton sector of our model, for parameter choices that are not in tension with $K$ and $D$ mixing constraints.

% \begin{figure}
%     \centering
%     \includegraphics[width=0.9\textwidth]{w12LFVtheta.pdf}
%     \caption{...}
%     \label{fig:w12LFVtheta}
% \end{figure}

% \textcolor{blue}{\subsection*{Anything else?}}
% \textcolor{blue}{\begin{itemize}
%     \item $K \to \pi \nu \nu$? NA62, maybe HIKE projection
% \end{itemize}}

The $W_{12}$ can additionally induce the LFV decay $\mu \to e \gamma$, via two loop running into the relevant dipole operators~\cite{Crivellin:2017rmk}. However, we find the numerical size of this effect is so small that the resulting constraints are unimportant compared to those from the tree-level $\mu \to 3 e$ decay.

\section{Phenomenology of the low mass $W_{23}$ triplet} \label{sec:W23}

The $W_{23}$ triplet has a richer phenomenology than $W_{12}$, not only due to its lower mass, but also due to its couplings to Higgs bosons. These arise because the $W_{23}$ contains a component of the $W_3$, which is the only gauge boson triplet in our model which couples to the Higgs. These couplings in turn modify the interactions of the Standard Model $W$ and $Z$ to fermions, via the SMEFT operators $O_{H l}^{(3)}$ and $O_{H q}^{(3)}$ (Eqs.~\eqref{eq:CHl3} and \eqref{eq:CHq3}). This makes electroweak constraints important, such as from $Z$ pole observables measured at LEP, see \S\ref{sec:ewFit} below.

The accidental $U(2)$ symmetry of the $W_{23}$'s coupling to the first two fermion generations ensures that there are no large effects in $K$ and $D$ flavour observables.\footnote{We remark that the phenomenology of heavy $SU(2)_L$ triplet vector bosons with accidental $U(2)$ flavour symmetries has been studied, in the context of flavour anomalies, in~\cite{Capdevila:2020rrl,Capdevila:2024gki}.
} Instead, the most pertinent quark flavour observables involve $b$ quarks, as discussed in the following subsection \ref{sec:quarkflavour}. Among lepton flavour observables, lepton flavour non-universality in $\tau$ decays is an important observable, as is $\mu \to 3 e$, which overcomes its $U(2)$ suppression due to being extremely well measured compared to LFV $\tau$ decays. Leptonic observables are discussed in \S\ref{sec:leptonic}.

Additionally, the $W_{23}$ can produce deviations in Drell--Yan processes at the LHC, due to its $O(1)$ couplings to first generation quarks. This is studied in \S\ref{sec:highPT}.

The $W_{23}$ provides a target for many upcoming measurements and experiments, including electroweak precision on the $Z$ pole at FCC-ee, new searches for $\mu \to 3 e$ at Mu3e, quark flavour experiments and Drell--Yan at colliders. These are discussed in \S\ref{sec:FCC}.

\subsection{Quark flavour observables} \label{sec:quarkflavour}

\subsubsection{Rare (semi-)leptonic $B_{(s)}$ decays}

The effective Lagrangian for leptonic and semileptonic $B$ and $B_s$ decays involving a $b\to s$ transition\footnote{Similar expressions hold for $b\to d$ transitions, up to appropriate relabelling.} can be written as
\begin{equation}
\mathcal{L}_{\text{eff}}=\frac{4 G_F}{\sqrt{2}}  \sum_{a} C_a O_a.
\end{equation}
The operators to which $W_{23}$ contributes at tree level are:
\begin{align}
    O_{9}^l&= V_{ts}^* V_{tb}\frac{e^2}{16\pi^2}\left(\bar s \gamma^\mu P_L b \right)  \left(\bar l \gamma_\mu l \right),\\
    O_{10}^l&= V_{ts}^* V_{tb}\frac{e^2}{16\pi^2}\left(\bar s\gamma^\mu P_L b \right)  \left(\bar l \gamma_\mu \gamma_5 l \right),  \\
    O_L^{\nu_l}&=V_{ts}^* V_{tb}\frac{e^2}{16\pi^2}(\bar s \gamma_\mu P_L b) (\bar{\nu}_{l} \gamma^\mu (1-\gamma_5) \nu_l),
\end{align}
where $l=\{e,\mu,\tau\}$,
and here we are focussing on lepton flavour conserving operators and assuming that $V_l\approx 1$ (we discuss departures from this assumption in the section on lepton flavour violation below).

Matching onto our model via the SMEFT, we find the BSM parts of the corresponding Wilson coefficients \cite{Aebischer:2015fzz,Jenkins:2017jig}:
\begin{align}
    C_9^e&=C_9^\mu =\frac{ \pi^2 v^2}{e^2 V_{tb}V^*_{ts}}\frac{1}{m_{23}^2}\frac{g_{SM}^2}{  s_\theta^2 c_\theta^2} \, [V_d]^*_{33}[V_d]_{32} (1-4 s_W^2 s_\theta^2),\\ 
    C_9^\tau&=-\frac{ 4\pi^2 v^2 s_W^2 }{e^2 V_{tb}V^*_{ts}}\frac{1}{m_{23}^2}\frac{g_{SM}^2}{  c_\theta^2 } \, [V_d]^*_{33}[V_d]_{32},\label{eq:C9tau}\\ 
    C_{10}^e &=C_{10}^\mu = -\frac{\pi^2 v^2}{e^2V_{tb}V^*_{ts}}\frac{1}{m_{23}^2}\frac{g^2_{SM}}{s_\theta^2 c_\theta^2}\,[V_d]^*_{33}[V_d]_{32}, \label{eq:C10eC10mu}\\
    C_{10}^\tau&=0,\label{eq:C10tau}\\
    C_L^{\nu_e}&=C_L^{\nu_\mu}=-\frac{ \pi^2 v^2}{e^2 V_{tb}V^*_{ts}}\frac{1}{m_{23}^2}\frac{g_{SM}^2}{  s_\theta^2 c_\theta^2} \,[V_d]^*_{33}[V_d]_{32},\\
    C_L^{\nu_\tau}&=0,
%    C_{V}^{eq}&=C_{V}^{\mu q}= \frac{v^2}{4 m_{23}^2}\frac{g_{SM}^2}{c_\theta^2s_\theta^2 }\,\left( c_\theta^2 (1+s_\theta^2) - \frac{1}{V_{qb}}\sum_k V_{u_k b} [V_u]^*_{3q}[V_u]_{3k}\right),\\
%    C_{V}^{\tau q}&= -\frac{v^2}{4 m_{23}^2}\frac{g_{SM}^2s_\theta^2}{c_\theta^2},
\end{align}
where $s_W^2 \approx 0.22$ is the squared sine of the Weinberg angle, and we have used the unitarity of the $V_d$ matrix, $\sum_i [V_d]^*_{i3}[V_d]_{i2}=0$. It is striking that $C_{10}^\tau$ and $C_L^{\nu_\tau}$ are both equal to zero, while their counterparts involving first and second generation leptons are not. This is due to the equality $C_{lq}^{(3)\tau\tau b s}=C_{Hq}^{(3)b s}$, which ultimately derives from the fact that the Higgs is treated as a third generation field, so has the same gauge coupling to the $W_{23}$ as the third generation leptons. The same cancellation occurs for the `deconstructed hypercharge' models of Refs.~\cite{Davighi:2023evx,FernandezNavarro:2023rhv}.

Note also that all of these Wilson coefficients depend on the choice for $V_d$: if we choose to take the `down-aligned' case in which $V_d=\mathbb{I}$, then $[V_d]_{32}=0$, and all of the Wilson coefficients are zero (and the model does not generate down-type FCNCs, as expected). On the other hand, in the `up-aligned' case with $V_d=V^\dagger$, the $V_d$ elements will cancel with the CKM elements in the denominator, and the Wilson coefficients will be independent of the CKM.

\paragraph{$\bm{B_s\to \mu\mu}$.} 
An important constraint on the model comes from the branching ratio of $B_s\to \mu\mu$, which is sensitive to the Wilson coefficient $C_{10}^\mu$. 
We take the combined measured value from \cite{Greljo:2022jac} (which combines the measurements of \cite{ATLAS-CONF-2020-049,LHCb:2021awg,CMS:2022mgd} following the methods of \cite{Aebischer:2019mlg,Altmannshofer:2021qrr}, and profiles over $B_d\to \mu \mu$):
\begin{equation}
BR(B_s\to \mu^+ \mu^-)=\left(3.32^{+0.32}_{-0.25} \right)\times 10^{-9},
\end{equation}
whereas the SM prediction is \cite{Altmannshofer:2021qrr}
\begin{equation}
BR(B_s\to \mu^+ \mu^-)_{SM}=\left(3.67\pm 0.15\right)\times 10^{-9}.
\end{equation}
The prediction for the branching ratio including the effects of BSM physics can be written as ({\em e.g.}~\cite{Allwicher:2023shc})
\begin{equation}
BR(B_s\to \mu^+ \mu^-)_{NP}=BR(B_s\to \mu^+ \mu^-)_{SM} \left|\frac{C_{10}^{\mu}+C_{10}^{SM}}{C_{10}^{SM}} \right|^2,
\end{equation}
where $C_{10}^{SM}(\mu_b)=-4.193\pm 0.033$ \cite{Blake:2016olu}.
The constraint on the NP part of the Wilson coefficient is then
\begin{align}
C_{10}^{\mu}\in [-0.20,0.53] ~~(2 \sigma).
\end{align}
The resulting constraints on the model are shown in Fig.~\ref{fig:bs23}, for two different values of $[V_d]_{32}$: the bounds shaded in lighter orange correspond to the up-aligned scenario in which $[V_d]_{32}^*=V_{cb}$, while the darker orange region is for an intermediate rotation with $[V_d]_{32}^*=V_{cb}/2$. In the down-aligned scenario (where $[V_d]_{32}^*=0$), there are no bounds at all. 

\paragraph{$\bm{B\to K^{(*)} \bar \nu\nu}$.}
The Belle II collaboration has recently measured a significant excess in the branching ratio of $B\to K \bar{\nu} \nu$, with respect to the SM prediction:
\begin{align}
    R^\nu_{K^{*}} &< 2.7 ~@ \,90\,\% ~\text{C.L.} \text{~\cite{Belle:2017oht}}\, , \label{eq:RKstarexp}\\
    R^\nu_K &= 2.8\pm 0.8 \text{~\cite{Belle-II:2021rof,Belle-II:2023esi}}\, , \label{eq:RKexp}
\end{align}
where the $R^\nu_K$ result is a combination of the new Belle II result with previous experimental limits~\cite{Greljo:2023bix}.
The branching ratios are \cite{Becirevic:2023aov,Buras:2014fpa}
\begin{equation}
    R^\nu_{K^{(*)}}\equiv \frac{BR(B\to K^{(*)} \bar \nu \nu)}{BR(B\to K^{(*)} \bar \nu \nu)_{\text{SM}}}= 1+ \sum_i \frac{2 \text{Re} [C_L^{SM} C_L^{NP, \nu_i \nu_i}]+|C_L^{NP, \nu_i \nu_i}|^2}{3|C_L^{SM}|^2},
\end{equation}
where $C_L^{\text{SM}}=-6.37(4)$ \cite{Buras:2014fpa,Altmannshofer:2009ma}. Since our model contributes only to the $C_L$ Wilson coefficient, it produces the same relative shift in $R^\nu_{K}$ and $R^\nu_{K^{*}}$. This is not the case for models that induce operators involving right-handed quarks~\cite{Buras:2014fpa,Becirevic:2023aov,Bause:2023mfe,Allwicher:2023syp}. 

As an indication of the favoured parameter space, in Fig.~\ref{fig:bs23} we show in green regions which lie within the $1\sigma$ range of \eqref{eq:RKexp}, while still in agreement with the $90\%$ C.L.~bounds of \eqref{eq:RKstarexp}. It is seen that in our model this favoured parameter space is always in conflict with bounds from $B_s\to \mu\mu$ and $B_s$ mixing. This can be understood from the fact that we have zero contribution to $C_L$ involving third generation leptons, so we cannot enhance $b\to s \nu \nu$ without also enhancing $b\to s \mu \mu$.

\subsubsection{$B_s$ meson mixing}\label{sec:Bs_mixing}
\begin{figure}
    \centering
    \includegraphics[width=\textwidth]{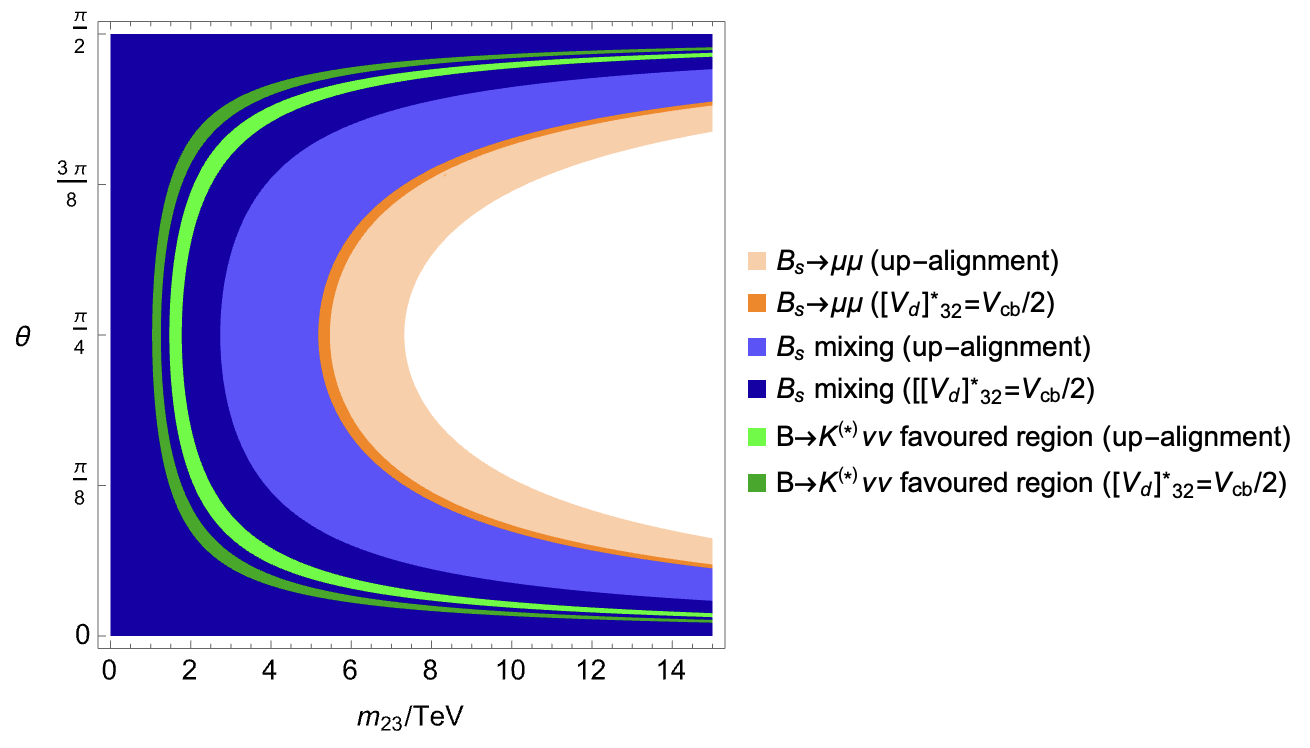}
    \caption{Constraints (95\% C.L.) on the parameter space of the $W_{23}$ gauge triplet from $B_s$ mixing and $B_s\to \mu\mu$ branching ratio. For each bound we show both the \emph{up-alignment} scenario -- with CKM misalignment between the down-quark mass basis and the gauge eigenbasis of the $W_{23}$ couplings -- and \emph{half-alignment}, with $[V_d]_{32}^*=V_{cb}/2$. Also shown are favoured regions that fit the observed deviation in $B\to K \bar\nu \nu$ at Belle II \eqref{eq:RKexp} (see text for how this region is defined). In our model this region is always well ruled out by other $b\to s$ constraints.}
    \label{fig:bs23}
\end{figure}

The $W_{23}$ can also contribute to $B_s$ mixing at tree level.
The effective Lagrangian for this process is:
\begin{equation}
\mathcal{L}_{\mathrm{eff}}\supset -C_1^{B_s} \left(\bar{b}_L \gamma^\mu s_L \right)^2.
\end{equation}
where in the SM:
\begin{align}
C_{1,SM}^{B_s}=\frac{G_F^2m_W^2}{4\pi^2}(V^*_{tb}V_{tq})^2S_0(x_t), ~~~~~~S_0(x_t)\approx 2.37,
\end{align}
while the $W_{23}$ contribution is
\begin{equation}
\label{eq:BsmixC1general}
C_1^{B_s}=\frac{1}{8m_{23}^2} \frac{g_{SM}^2}{s_\theta^2 c_\theta^2 } \left([V_d]_{33}[V_d]_{32}^* \right)^2.
\end{equation}
The measured value of $\Delta M_s$ and the weighted average of its theory predictions are respectively:
\begin{align}
%\Delta M_d^{exp}&=(0.5064\pm 0.0019) \mathrm{ps}^{-1}\\
\Delta M_s&=(17.7656\pm 0.0057)\, \mathrm{ps}^{-1}\text{~\cite{LHCb:2021moh}}\, ,\\
\Delta M_s^{SM}&=(18.4^{+0.7}_{-1.2})\, \mathrm{ps}^{-1} \text{~ \cite{DiLuzio:2019jyq}}\, ,
\end{align}
from which the NP contribution to $\Delta M_s^{SM}$ is then bounded to be
\begin{equation}
\delta(\Delta M_s)=\frac{\Delta M_s-\Delta M_s^{SM}}{\Delta M_s^{SM}}\in [-0.11,0.095] ~~\,(95 \%)\, ,
\end{equation}
where
\begin{equation}
\delta(\Delta M_s)=\left|1+\frac{C_{1,NP}^{B_s}}{C_{1,SM}^{B_s}} \right|-1\, .
\end{equation}
In the down-aligned scenario, this imposes no bounds on the model. The bounds for fully up-aligned, and for an intermediate situation in which $[V_d]_{32}^*=V_{cb}/2$, are shown in Fig~\ref{fig:bs23}.
The shapes of the $B_s\to \mu\mu$ bounds and the $B_s$ mixing bounds are similar, because of the identical parametric dependence of the Wilson coefficients in eqs.~\eqref{eq:BsmixC1general} and \eqref{eq:C10eC10mu}. However, the $B_s\to \mu\mu$ constraints are systematically stronger in the scenarios considered. If the $V_d$ rotation were very large (such that $[V_d]_{32}^*\gg V_{cb}$) then the $B_s$ mixing constraints would eventually become dominant, due to the quadratic dependence of $C_1^{B_s}$ on $[V_d]_{32}^*$, whereas $C_{10}^\mu$ has only a linear dependence.

\subsubsection{Comments on charged currents and $R_D^{(*)}$}

The charged components of $W_{23}$, being heavy flavour non-universal counterparts of the $W^\pm$ bosons, will induce effects in charged current decays such as $B\to \mu \nu$. But we find these observables to be much less sensitive than others we consider: if we are not in the down-aligned case, then the $Z_{23}$-induced FCNCs are much more important, since the competing SM processes are loop- and GIM-suppressed. Meanwhile, even if we are in the down-aligned case, then electroweak or leptonic observables are more sensitive (see following subsections). 

Since $W_{23}$ couples lepton non-universally, one might expect it to produce sizeable effects in $R_D^{(*)}$, which currently show a $3.3\sigma$ deviation from SM predictions~\cite{HeavyFlavorAveragingGroup:2022wzx}. However, unlike models that successfully explain this deviation via couplings which are broadly aligned to the third generation of quarks and leptons, $W_{23}$ instead has couplings to first and second generation leptons which are strictly greater than $g_{SM}$. This inability to decouple from the light generations also pushes the model to higher mass than is needed to fit the anomaly in $R_D^{(*)}$: as we will see in the combination plot Fig.~\ref{fig:W23_summary_current}, $m_{23}$ must be at least $9$ TeV given current measurements. This already exceeds the perturbative unitarity bound for vector states explaining the $R_D^{(*)}$ anomaly~\cite{DiLuzio:2017chi}.

\subsection{Lepton flavour observables}\label{sec:leptonic}

We now consider the constraints coming from the lepton sector. In general these are weaker than the constraints coming from quark flavour observables -- unless of course the mixing angles that give rise to charged lepton flavour violation (cLFV) are large, which is not expected to be the case in our flavour model. We begin in \S \ref{sec:tau} with `flavour-conserving' constraints that probe lepton flavour universality violation (LFUV) in tau decays. This is essentially independent of the lepton mixing matrix, and can be computed to good approximation just using the `gauge eigenbasis' leptonic couplings of the $W_{23}$ triplet. Of course, the cLFV bounds that we compute in \S \ref{sec:cLFV} depend on the mixing angles, which we parametrize via a real rotation.

\subsubsection{LFUV tests in tau decays} \label{sec:tau}

We consider the following set of observables that probe LFUV (tau {\em vs} light leptons) in leptonic and hadronic tau decays:
\begin{align}
    \frac{g_\tau}{g_{\ell}} &= \frac{BR(\tau \to \ell \nu \bar{\nu})/BR(\tau \to \ell \nu \bar{\nu})_{\text{SM}}}{BR(\mu \to e \nu \bar{\nu})/BR(\mu \to e \nu \bar{\nu})_{\text{SM}}}, \qquad \ell \in \{e,\mu\}, \label{eq:tauLFUVleptonic}\\
    \left.\frac{g_\tau}{g_{\ell}}\right|_{H} &= \frac{BR(\tau \to H \nu )/BR(\tau \to H \nu)_{\text{SM}}}{BR(H \to \mu\bar{\nu})/BR(H \to \mu \bar{\nu})_{\text{SM}}}, \qquad H \in \{\pi, K\}\, .\label{eq:tauLFUVhadronic}
\end{align}
These observables are constructed such that they equal unity in the SM. The experimental measurements (coming from Belle,\footnote{
Recently, the Belle II collaboration announced a new measurement of the $g_\mu/g_e$ LFUV ratio in tau decays~\cite{BelleIItau}, using $362 \text{~fb}^{-1}$ of data. This is the most precise single measurement of $\mu$ {\em vs.}~$e$ LFUV in tau decays, but since the $W_{23}$ triplet of our model couples universally to electrons and muons this does not provide an important constraint on our flavour model. The contributions from the $W_{12}$ are at too high a scale to give any relevant deviation.
} amongst others) probe these observables to part-per-mille precision, and are consistent with unity at the $2\sigma$ level - see the HFLAV combinations (including correlations) in Ref.~\cite{HeavyFlavorAveragingGroup:2022wzx}, which we use here. 

In the presence of LFUV new physics, such as is generated by integrating out the $W_{23}$ triplet in our model (which couples differently to tau {\em vs} light leptons), these observables will deviate from unity and so be constrained -- for example, Ref.~\cite{Allwicher:2021ndi} explored how these observables constrain leptoquark models designed to explain anomalies in $B$-meson decays. 
We use the SMEFT expressions for these observables given in~\cite[Appendix A.1]{Allwicher:2023shc}. 
In terms of our model parameters, we have
\begin{equation}
    \frac{g_\tau}{g_e}=\frac{g_\tau}{g_\mu}=1-\frac{v^2 g_{SM}^2}{4m_{23}^2 s^2_\theta c_\theta^2}
\end{equation}
for the purely leptonic ratios, and 
\begin{align}
    \left.\frac{g_\tau}{g_{\ell}}\right|_{\pi} &\approx 1-\frac{v^2 g_{SM}^2}{4m_{23}^2 s_\theta^2 c_\theta^2} \left( 1 + \mathcal{O}(\lambda^6)
    \right) \, , \\
    \left.\frac{g_\tau}{g_{\ell}}\right|_{K} &\approx 1-\frac{v^2 g_{SM}^2}{4m_{23}^2 s_\theta^2 c_\theta^2} \left( 1 + \mathcal{O}(\lambda^5)
    \right)
\end{align}
for the hadronic observables, where $\lambda$ is the Cabibbo angle. We notice that the model predicts a universal shift in all the tau LFUV ratio observables, up to small non-universal corrections suppressed by quark mixing angles in the case of the hadronic observables.
By computing the $\Delta \chi^2$ statistic as a function of our model parameters $(m_{23},\theta)$, we obtain the 95\% C.L. contours plotted in Fig.~\ref{fig:W23LFV}.

%%%%%%%%%%%
\subsubsection{Charged lepton flavour violation} \label{sec:cLFV}

To investigate the charged lepton flavour violation that can be induced by $W_{23}$, we can again use the parameterisation of the $V_l$ mixing matrix given in Eq.~\eqref{eq:Vlrotsmallangles}. To second order in the mixing angles, this results in a leptonic coupling matrix of the form
\begin{equation}
g^l_{[23]}=\frac{g_{SM}}{s_\theta c_\theta}\begin{pmatrix}
c_\theta^2 - \beta^2 & \beta \gamma & \beta \\
\beta \gamma & c_\theta^2 - \gamma^2 & -\gamma \\
\beta & - \gamma & - s_\theta^2 + \beta^2 + \gamma^2
\end{pmatrix}.
\end{equation}
Recall from \S \ref{sec:12_LFV} that, while the charged lepton mixing matrix is in general unitary rather than orthogonal, because the branching ratios that constrain lepton flavour violation are insensitive to phases we can approximate the mixing as real. 
Notice that (due to the $U(2)$ symmetry of the $W_{23}$ couplings) the off-diagonal couplings involving a $\tau$ are induced at linear order in the angles $\gamma$ and $\beta$, while the off-diagonal couplings in the first two generations are proportional to $\beta \gamma$. Nevertheless, the much higher precision on the experimental bounds of $\mu \to e$ processes compared to $\tau \to (\mu, e)$ processes can overcome this suppression to impose important constraints on $W_{23}$ parameter space, as we shall see in the following.

\begin{figure}
    \centering
    \includegraphics[height=0.5\textwidth]{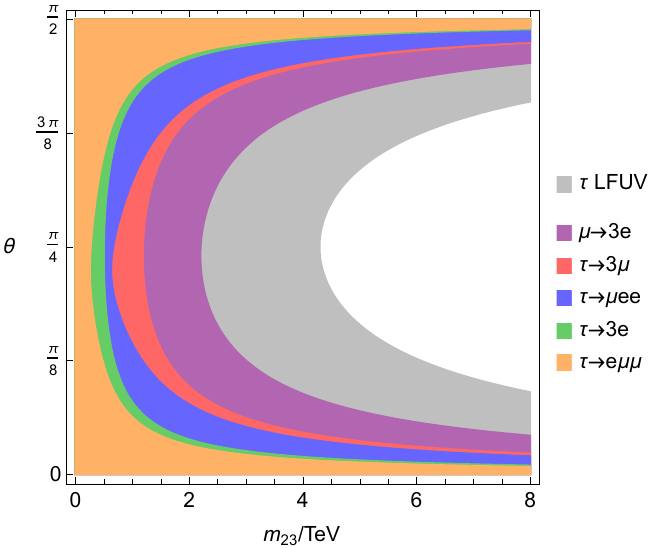}
    \caption{Constraints (95\% C.L.) from measurements of lepton flavour violating and lepton flavour non-universal $\tau$ and $\mu$ decays on the parameter space of $W_{23}$, for CKM-like charged lepton mixing angles $\gamma=0.05$, $\beta=0.01$. }
    \label{fig:W23LFV}
\end{figure}

We use the formulae of \cite{Crivellin:2013hpa} for the branching ratios in terms of the SMEFT coefficients (\S\ref{sec:SMEFT}), to find:
\begin{align}
BR(\tau \to 3 \mu)
& =  \frac{m_\tau^5}{3072 \pi^3 \Gamma_\tau}\left(\frac{\gamma g_{SM}^2}{4 m_{23}^2 c_\theta^2s_\theta^2}\right)^2 \bigg(8 s_W^4 s_\theta^4 +\left(c_\theta^2+2 (1-2s_W^2)s_\theta^2\right)^2 \bigg), \\
BR(\tau \to 3 e)
& =  \frac{m_\tau^5}{3072 \pi^3 \Gamma_\tau}\left(\frac{\beta g_{SM}^2}{4 m_{23}^2 c_\theta^2s_\theta^2}\right)^2 \bigg(8 s_W^4 s_\theta^4 +\left(c_\theta^2+2 (1-2s_W^2)s_\theta^2\right)^2 \bigg), \\
BR(\tau \to \mu ee) & =  \frac{m_\tau^5}{6144 \pi^3 \Gamma_\tau}\left(\frac{\gamma g_{SM}^2}{4 m_{23}^2 c_\theta^2s_\theta^2}\right)^2 \bigg(16 s_W^4 s_\theta^4 +\left(c_\theta^2+2 (1-2s_W^2)s_\theta^2\right)^2 \bigg),\\
BR(\tau \to e \mu \mu) & =  \frac{m_\tau^5}{6144 \pi^3 \Gamma_\tau}\left(\frac{\beta g_{SM}^2}{4 m_{23}^2 c_\theta^2s_\theta^2}\right)^2 \bigg(16 s_W^4 s_\theta^4 +\left(c_\theta^2+2 (1-2s_W^2)s_\theta^2\right)^2 \bigg),\\
BR(\mu \to 3 e) & =  \frac{m_\mu^5}{3072 \pi^3 \Gamma_\mu}\left(\frac{\gamma \beta g_{SM}^2}{4 m_{23}^2 c_\theta^2s_\theta^2}\right)^2 \bigg(8 s_W^4 s_\theta^4 +\left(c_\theta^2+2 (1-2s_W^2)s_\theta^2\right)^2 \bigg).
\end{align}
The most natural expectation within the model, which treats quarks and leptons democratically, is for the lepton mixing matrix $V_l$ to be `CKM-like', {\em i.e.} with similar size mixing angles to the $V_{u,d}$ matrices. In this scenario we expect $\beta \sim 0.01$, $\gamma \sim 0.05$. We use these values to calculate the bounds in Fig.~\ref{fig:W23LFV}, where the experimental bounds on $\tau$ decays are from the Belle experiment~\cite{Hayasaka:2010np}, and the bound on $BR(\mu\to 3 e)$ is from the SINDRUM experiment~\cite{SINDRUM:1987nra}.

The constraint from $\mu \to 3 e$ is stronger than that from any $\tau$ decay process,\footnote{As was the case for the $W_{12}$ discussed previously in \S \ref{sec:12_LFV}, the $W_{23}$ can additionally induce the decay $\mu \to e \gamma$ via two loop running into the relevant dipole operators~\cite{Crivellin:2017rmk}. Again, the numerical size of this effect is so small that the resulting constraints are unimportant compared to the tree-level decays above.}
since the suppression due to the $U(2)$ flavour symmetry is compensated for by the higher precision on the measurement. Nevertheless, the cLFV bounds are much less constraining than the LFUV bounds discussed in the previous subsection, meaning that CKM-like mixing angles are allowed in the model given other constraints.

\subsection{Electroweak observables} \label{sec:ewFit}

Due to its coupling to Higgs doublets, the $W_{23}$ induces significant effects in electroweak (EW) precision observables such as those measured on the $Z$ pole at LEP. In particular, the $O_{H l}^{(3)}$ and $O_{H q}^{(3)}$ SMEFT operators, whose Wilson coefficients are given in Eqs.~\eqref{eq:CHl3} and \eqref{eq:CHq3}, modify SM gauge boson couplings to leptons and quarks respectively.

There are also important indirect effects due to SMEFT operators that modify muon decay and therefore shift the extracted value of $G_F$: in our model the non-zero operators are $(O_{ll})_{1221}$, $(O_{H l}^{(3)})_{11}$ and $(O_{H l}^{(3)})_{22}$. Following~\cite{Berthier:2015oma}, we take as a set of input parameters the $Z$ pole mass, Fermi constant and electromagnetic fine structure constant, {\em i.e.} the triple $\{ m_Z, G_F, \alpha_\text{e} \}$, meaning that the operators entering this shifted value of $G_F$ propagate indirectly into every electroweak precision observable. 

In the following subsections we discuss some important observables for our model, and then present a fit to a full set of electroweak precision constraints.

\subsubsection{$W$ mass} \label{subsubsec:Wpole}

In the presence of dimension 6 operators the deviation of the predicted value of the $W$ pole mass is given by \cite{Bjorn:2016zlr}
\begin{align}
    \delta m_W^2 &= (m_W^2)_\text{SM} - (m_W^2)_\text{SMEFT} \\
                &= (m_W^2)_\text{SM} \,\Delta \left[ 4\, C_{HWB} + \frac{c_W}{s_W} C_{HD} + 2\frac{s_W}{c_W}(C_{Hl}^{(3)11}+C_{Hl}^{(3)22} -  C_{ll}^{1221}) \right],
\end{align}
with $\Delta = 2\sqrt{2}G_F c_Ws_W/(c^2_W- s^2_W)$, where $s_W$ ($c_W$) is the sine (cosine) of the Weinberg angle.
In the EFT of the $W_{23}$, the Wilson coefficients that are non-zero at tree level are $C_{Hl}^{(3)11}$, $C_{Hl}^{(3)22}$ and $C_{ll}$. When written in terms of the model parameters (neglecting the small contributions from the charged lepton mixing angles $\beta$ and $\gamma$) the shift in the $W$ mass is given as
\begin{equation}\label{eq:wPoleShift}
     \delta m_W^2 = (m_W^2)_\text{SM}\,\Delta\, \frac{s_W}{c_W} g^2_\text{SM} \left(\frac{2 + \cot^2 \theta}{2m^2_{23}}\right) .
\end{equation}
Notice that (\ref{eq:wPoleShift}) is strictly positive and as a result the predicted value from the EFT of the $W$ pole mass is always less than the SM prediction: $ (m_W^2)_\text{SMEFT} < (m_W^2)_\text{SM}$. 

Recent experimental determinations of the $W$ mass have found a central value {\em greater} than the SM prediction. In particular, prior to the recent CDF 2022 determination~\cite{CDF:2022hxs}, the PDG average was $m_W=80.377 \pm 0.012$ GeV~\cite{Workman:2022ynf}, which is mainly determined by Tevatron 2013~\cite{CDF:2013dpa} and ATLAS 2017~\cite{ATLAS:2017rzl} measurements. This shows a nearly 2$\sigma$ deviation with respect to the SM prediction~\cite{Awramik:2003rn}. Since the contribution of the $W_{23}$ only reduces the mass and therefore increases the tension, the $W$ mass provides a strong constraint on the model parameters. We do not use the newer CDF 2022 measurement in our analysis here, since it is in disagreement both with the SM and with other $m_W$ measurements. Including it in the fit would increase the tension both with the SM and with our model.

\subsubsection{$Z$-pole fermion asymmetries: $A_e$ and $A_{FB}^b$} \label{subsubsec:AFB_Zee}

The parity structure of the $Z$ boson couplings to fermions was measured precisely at SLD and LEP through the fermion asymmetry observables $A_f$ and $A_{FB}^f$~\cite{ALEPH:2005ab}, which are defined as
\begin{align}
  A_f = \frac{2 g_V^f g_A^f}{(g_V^f)^2 + (g_A^f)^2} = \frac{(g_L^f)^2-(g_R^f)^2}{(g_L^f)^2+(g_R^f)^2},  \qquad A_{FB}^f = \frac{3}{4} A_l A_f.
\end{align}
In the SMEFT, the vectorial and axial couplings $g^{f}_{V,A}$ of the $Z$ to fermions $f$ receive corrections $g^{f}_{V,A}=g^{f,SM}_{V,A}+\delta g^{f}_{V,A}$ \cite{Brivio:2017vri} %\textcolor{red}{Comment 7: Sufficient to add the expressions for the corrections in terms of WCs?}. 
For both down-type quarks and charged leptons, the contribution of $W_{23}$ produces \emph{positive} contributions to both the vectorial and axial shifts. Explicitly, the shifts in the couplings to electrons and $b$ quarks are given by
\begin{align}
    \delta g^e_{V}&= \frac{1}{16\sqrt{2}G_F m_{23}^2}\frac{g_{SM}^2}{(g_{SM}^2-g_Y^2)}\left(4 g_{Y}^2+\frac{1}{2}( g_{SM}^2+3g_Y^2) \cot^2 \theta \right),\\
    \delta g^e_{A}&= \frac{1}{32\sqrt{2}G_F m_{23}^2}g_{SM}^2  \cot^2 \theta,\\
    \delta g^b_{V}&= \frac{1}{96\sqrt{2}G_F m_{23}^2}g_{SM}^2\left(6\tan^2 \theta + \frac{(3g_{SM}^4-4 g_L^2g_Y^2 + 9g_Y^4)}{(g_{SM}^4-g_Y^4)}\left(2+ \cot^2 \theta\right)\right),\\
    \delta g^b_{A}&= \frac{1}{32\sqrt{2}G_F m_{23}^2}g_{SM}^2\left(2\tan^2\theta + \frac{1}{2}\frac{(g_{SM}^2-3g_Y^2)}{(g_{SM}^2+g_{Y}^2)}\left(2+ \cot^2 \theta \right)\right),
\end{align}
where $g_Y$ is the hypercharge gauge coupling. These shifts are always positive, for any value of $\theta$, and the magnitude of the vectorial shift is always larger than that of the corresponding axial shift.
To leading order in $1/m_{23}^2$, the contributions to $A_e$, $A_b$ and $A_{FB}^b$ are given in terms of these as:
\begin{align}
 \delta A_e&=-\frac{16 g_Y^2 (g_{SM}^4-g_Y^4)}{(g_{SM}^4-2g_{SM}^2 g_Y^2+5 g_Y^4)}\left( g_{SM}^2 (\delta g_V^e-\delta g_A^e)+g_Y^2 (\delta g_V^e+3\delta g_A^e)\right),\\
 \delta A_b &=  -\frac{48 g_Y^2(g_{SM}^2+g_Y^2)(3g_{SM}^2+g_Y^2)}{(9g_{SM}^4+6g_{SM}^2 g_Y^2+5 g_Y^4)}\left( 3g_{SM}^2 (\delta g_V^b-\delta g_A^b)+g_Y^2 (3\delta g_V^b+\delta g_A^b)\right),\\
 \delta A_{FB}^b&= \frac{3}{4}\left(A_e^{SM} \,\delta A_b + A_b^{SM} \,\delta A_e\right).
\end{align}
Given that $\delta g_V^{e,b}>\delta g_A^{e,b}$ and $A_{e,b}^{SM}>0$, it can be seen that the shifts $\delta A_e$, $\delta A_b$ and $\delta A_{FB}^b$ are each \emph{negative} in our model.

The asymmetry $A_e$ was measured at SLD and LEP~\cite{ALEPH:2005ab}, and the combination has a small tension with respect to the SM in the positive direction. Since $W_{23}$ reduces $A_e$ with respect to the SM, it can only increase the tension and this observable therefore places quite a strong constraint on the model.

The forward-backward asymmetry in the production of $b$ quark pairs, $A_{FB}^b$ also shows a tension between the SM prediction and its measured value~\cite{ALEPH:2005ab}. In this case, the deviation requires a reduction in the prediction for $A_{FB}^b$, so the contribution from $m_{23}$ moves the prediction in the preferred direction. But to achieve the experimental central value, the parameters of the model must be within a region already well ruled out in the overall fit, requiring a mass of the order of $2$ TeV for $\theta \sim \pi/4$.

\subsubsection{Fit to electroweak precision observables} \label{subsubsec:EWfit}

\begin{figure}[t]
    \centering
    \includegraphics[width=0.7\textwidth]{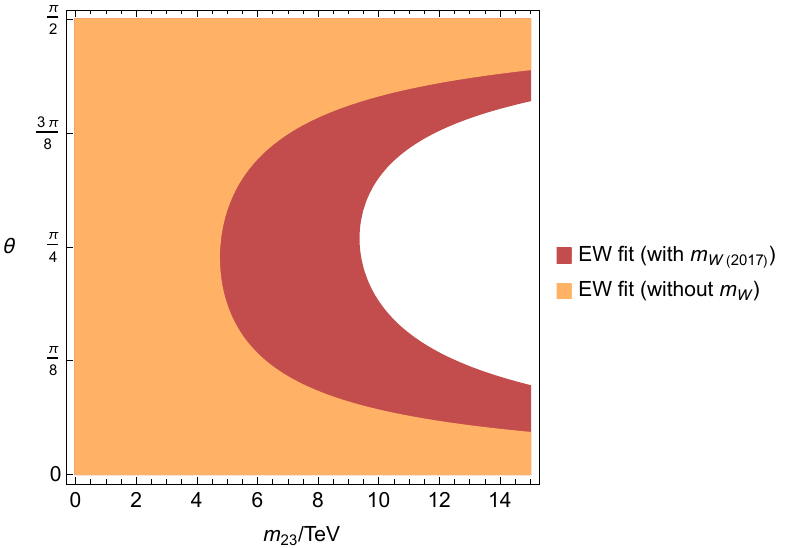}
    \caption{Constraints (at 95\% C.L.) on $m_{23}$ and $\theta$ from EW observables. The dark red region is excluded by a fit including electroweak precision observables on the $Z$ and $W$ poles, and the measured value of $m_W$ from ATLAS 2017~\cite{ATLAS:2017rzl} and Tevatron 2013~\cite{CDF:2013dpa}. The orange region is excluded by the same fit without the inclusion of the $m_W$ observable. 
    %Included are the global constraints from all EW observables and also the most constraining observables: $m_W$, $\Gamma_Z$, $\text{A}(Z \rightarrow ee)$ and $R_\mu$.
    }
    \label{fig:ewConstraints}
\end{figure}

The results of a full fit to electroweak precision observables is shown in Fig.~\ref{fig:ewConstraints}. The fit uses the likelihood of Ref.~\cite{Breso-Pla:2021qoe}, which includes $Z$ and $W$ pole observables from LEP, SLD and the Tevatron, as well as $W$ branching fractions as Drell--Yan forward-backward asymmetries at the $Z$ peak from the LHC. We also performed a fit using the \texttt{smelli}~\cite{Aebischer:2018iyb,Straub:2018kue,Aebischer:2018bkb} package as a cross-check, finding good agreement.

The orange region in Fig.~\ref{fig:ewConstraints} shows the $2\sigma$ excluded region from the fit excluding the $m_W$ observable, while in dark red is the excluded region including $m_W$ (using Tevatron 2013~\cite{CDF:2013dpa} and ATLAS 2017~\cite{ATLAS:2017rzl} measurements). As discussed in \S\ref{subsubsec:Wpole} above, the $W$ mass observable provides a strong constraint on the $W_{23}$ parameter space (nearly doubling the excluded values of $m_{23}$), since it is already close to $2\sigma$ in tension with the SM, and the contribution of the $W_{23}$ can only increase the tension. We do not include the most recent CDF measurement~\cite{CDF:2022hxs} of $m_W$ in our analysis,\footnote{Nor do we include the LHCb 2021 measurement~\cite{LHCb:2021bjt} of the $W$ mass, which (for now) has significantly larger uncertainty than the measurements we include.} which would increase the tension both with the SM and with our model. 

In later combination fits, we show the EW constraint including the $m_{W(2017)}$ measurements, however it should be kept in mind that new and future measurements of $m_W$ can change the exclusions significantly.

\subsection{High $p_T$ Drell--Yan observables} \label{sec:highPT}

For the $W_{23}$ triplet of gauge bosons, which do not cause flavour violation in the 1-2 sector and so are permitted (by flavour constraints) to be relatively light, there are relevant constraints coming from high-mass searches at the LHC. We here compute the constraints from high-$p_T$ Drell--Yan tails using the \texttt{HighPT} Mathematica package~\cite{Allwicher:2022mcg,Allwicher:2022gkm}, from ATLAS and CMS searches in $pp \to \ell\ell$ and $pp \to \ell\nu$ channels, for all three lepton flavours.

%\paragraph{SMEFT operators.}
The important sector of SMEFT operators for these observables is those 4-fermion operators (\ref{eq:Clq}) with two quark and two lepton fields. For high-$p_T$ bounds, in contrast to flavour bounds, we can approximate the flavour structure as being diagonal (because the flavour off-diagonal elements are small), and essentially set $V_u \approx 1 \approx V_l$ when computing the bounds in this Subsection. The relevant Wilson coefficient contributions coming from the $W_{23}$ triplet are then simple functions of the model parameters $(m_{23}, \theta)$:
\begin{align}
    (C_{lq}^{(3)})_{\alpha\alpha\beta\beta}&= -\frac{g_{SM}^2 \cot^2\theta}{4m_{23}^2}, \\
    (C_{lq}^{(3)})_{\alpha\alpha 33} = (C_{lq}^{(3)})_{33\alpha\alpha} &= \frac{g_{SM}^2}{4m_{23}^2}, \label{eq:highPT-WC-heavy-light}\\
    (C_{lq}^{(3)})_{3333}&= -\frac{g_{SM}^2 \tan^2\theta}{4m_{23}^2} \, ,
\end{align}
where $\alpha,\, \beta = 1,\, 2$ denote light family indices. Note that, as always, these $W_{23}$-induced Wilson coefficients are independent of the second angular parameter $\phi = \tan^{-1}(g_2/g_1)$, which measures the 1-2 flavour universality breaking effects (coming only from the $W_{12}$ triplet, that is responsible for resolving the 1-2 flavour structure at higher scales).

%\paragraph{Code setup.}
We use the \texttt{HighPT} package~\cite{Allwicher:2022mcg} to compute the likelihood function of our model parameters, $\chi^2(m_{23},\theta)$,
given all the relevant ATLAS and CMS searches in both $\ell\ell$~\cite{ATLAS:2020zms,CMS:2021ctt} and $\ell\nu$~\cite{ATLAS:2019lsy,ATLAS-CONF-2021-025} channels -- a total of six searches, all of which use $139~\text{fb}^{-1}$ or thereabouts.
We use \texttt{HighPT} in its `SMEFT mode', which means that the expected yield in each bin is calculated in the model assuming 4-point contact SMEFT interactions, rather than explicitly modelling the kinematic effects of the mediators. We do this for practical reasons,
since the mediator mode is not yet implemented for $W'$ and $Z'$ gauge boson mediators; we nonetheless expect this to be a decent approximation, because the searches constrain us to mass regimes larger than 4-5 TeV which is near (or beyond) the tail of the search, and so even for the $s$-channel $Z'$ the SMEFT approximation should not be too bad. 
From this likelihood we obtain 95\% C.L. constraints on our 2-parameter plane by plotting $\Delta \chi^2 = \chi^2 - \chi^2_{\mathrm{min}}= 5.99$ contours.

\begin{figure}
    \centering    \includegraphics[width=0.83\textwidth]{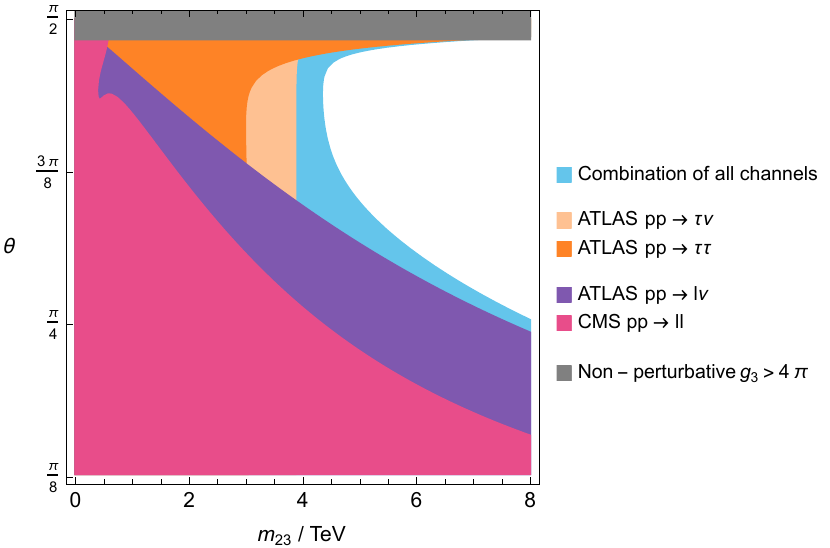}
    \caption{High $p_T$ bounds on the $W_{23}$ triplet states, coming from $pp \to \ell\ell$ and $pp \to \ell \nu$ Drell--Yan measurements at the LHC for all lepton flavours, computed using the \texttt{HighPT} package~\cite{Allwicher:2022mcg}. The white region is currently preferred at 95\% CL. }
    \label{fig:highpt}
\end{figure}

From the combination of all channels (the light blue region in Fig.~\ref{fig:highpt}), we find the current limit on the mass is
\begin{equation} \label{eq:M23-highpT-bound}
    m_{23} \geq 4.4 \text{~TeV}.
\end{equation}
This lightest permissible mass is achieved for $\theta$ values roughly in the range $[1.35,1.45]$, {\em i.e.}
    $4 \lesssim \frac{g_3}{\sqrt{g_1^2+g_2^2}} \lesssim 8$.
That is, the current direct search bound is weakest for new physics states that are coupled more strongly to the third family than the light families. This is typically what one would expect from any direct search bound coming from hadronic colliders, since production is usually driven by the light-quark interactions for which the PDFs are enhanced in $pp$ collisions.

Let us understand why this parameter space is preferred, physically, in a little more detail.
For $\theta$ close-ish to $\pi/2$,
we minimize couplings to the light generations and so avoid the strong bound from the single muon channel. But one cannot live in parameter space too close to $\pi/2$, for which $g_3$ is blowing up, because then the mono-tau and di-tau production accordingly diverges (the orange bands in Fig.~\ref{fig:highpt}). More precisely, this divergence is driven by $b\bar{b}$ production via the $(C_{lq}^{(3)})_{3333}$ Wilson coefficient, for which the divergent parton-level cross-section eventually overcomes the PDF suppression. Note that these mono-tau and di-tau searches exclude the `non-perturbative' region described above, which is indicated by the dark grey band, where the gauge coupling $g_3 = g_{SM}/\cos\theta$ exceeds $4\pi$.\footnote{The point $\theta=\pi/2$ corresponds to the limit $g_{1,2}/g_3 \to 0$. Recalling $g_{1,2} \geq g_{SM}$, this clearly violates perturbativity. We can estimate a perturbative limit by requiring
\begin{equation}
    g_3 \leq 4\pi \implies |\theta-\pi/2| \geq |\arccos(g_{SM}/4\pi)|\approx 0.05\, ,
\end{equation}
using $g_{SM}=0.64$. 
This translates to requiring $g_3/\sqrt{g_1^2+g_2^2} < 20$ or so. } 
Also notice that the tau constraints do not decouple in the converse limit where $\theta \to 0$. This is because there is a cancellation of the $\theta$-dependence in the mixed heavy-light flavour semi-leptonic WCs (\ref{eq:highPT-WC-heavy-light}), so that the contact interaction involving two tau-flavoured leptons plus two valence quarks is actually independent of $\theta$. 

Three further important conclusions we can draw from Fig.~\ref{fig:highpt} are:
\begin{itemize}
    \item The di- and mono-tau searches are the most important in setting the lower-bound (\ref{eq:M23-highpT-bound}) on the mass $m_{23}$ of the $SU(2)_L$ triplet $W_{23}$. Without these tau flavoured LHC searches, $m_{23}$ is not even ruled out at 1 TeV (for large $g_3$) by high-energy observables.
    \item On the other hand, the muon searches (blue) dictate the $\theta$-dependence of the bounds at large masses ($\gtrsim 5$ TeV); in particular, the muon searches rule out parameter space with large $g_{1,2}/g_3$ (regions with $\theta$ approaching $0$).
    \item For muonic channels (and for tau, away from the $g_3 \gg g_{SM}$ limit), the single lepton channels give stronger constraints than the di-lepton, for our $SU(2)_L$-based models in which couplings to $\ell\nu$ pairs are the same as couplings to $\ell\ell$ by $SU(2)_L$ symmetry. This traces back to the strength of the experimental searches, which is higher for $l\nu$. We refer the reader to~\cite[Figure 4]{CMS:2021ctt} (for $ll$) {\em vs.}~\cite[Figure 2]{ATLAS:2019lsy} (for $l\nu$) for the experimental plots corresponding to the data used in our \texttt{HighPT} computation. This difference is largely because the background is simpler for the charged-current $l\nu$ channels, which is dominated by Drell--Yan production of $W$ bosons (for the $ll$ channels there are more backgrounds, including photon-induced processes as well as $Z$ and Higgs).
\end{itemize}
When high-$p_T$ LHC data is taken in isolation, we have learnt that searches in $\tau$ final states are important in setting the allowed mass range for the $W_{23}$ triplet. But the low-mass parameter space region allowed by  high-$p_T$ -- around 5 TeV, and for angles within about $(90\pm 15)^{\circ}$ -- is excluded at 95\% CL by EWPOs, as in Fig.~\ref{fig:ewConstraints}. 
So, when collider and electroweak constraints are taken together (see \S \ref{sec:combination}), it is actually the {\em single muon} LHC search constraint that cuts important parameter space otherwise permitted by electroweak precision. See \S \ref{sec:combination} for a discussion of the parameter space accounting for all the complementary constraints.

Lastly, for completeness we record that one can also compute similar high-$p_T$ bounds on the $W_{12}$ triplet. We find a quantitatively similar bound on the mass of $m_{12} \geq 4.5$ TeV or so, this time driven by the mono-muon (followed by mono-electron) searches. Because the flavour bounds, particularly from meson mixing (Fig.~\ref{fig:KandDM12}) already constrain the $W_{12}$ states to be much heavier than this (100s of TeV), we conclude that high-$p_T$ does not provide a relevant constraint on these high-mass states; high-$p_T$ observables are only relevant for the $W_{23}$ bosons.

\subsection{Combination of constraints} \label{sec:combination}

We have learnt that there are important and complementary constraints on the parameter space of the light $W_{23}$ triplet, that arises from deconstructing the electroweak $SU(2)_L$ gauge symmetry into $SU(2)_{L,1+2}\times SU(2)_{L,3}$. We now put things together to delineate the viable parameter space of the model.

\begin{figure}
    \centering    
    \includegraphics[width=0.9\textwidth]{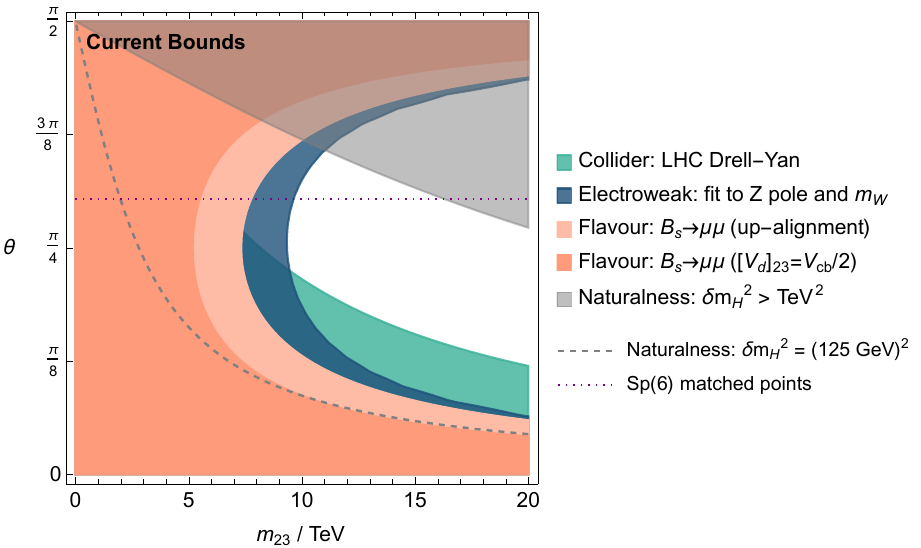}
    \caption{
    Summary plot showing the current most important constraints on the $(m_{23},\theta)$ parameter space of the $W_{23}$ triplet that arises in the flavour-deconstructed $SU(2)_L$ model.
    We see the constraints coming from colliders, electroweak precision observables (EWPOs), and flavour are highly complementary. The most constraining flavour observable is $BR(B_s \to \mu^+ \mu^-)$, which we plot here under two different flavour alignment scenarios (orange). The stronger version of the flavour bound, corresponding to $V_{cb}$ coming purely from the down-type mixing, is still weaker than the bound from EWPOs (dark blue), which excludes $m_{23}$ lighter than 9.4 TeV. The collider bounds (green), which come from Drell-Yan $pp \to \ell \ell(\nu)$ for all possible final state lepton flavours, exclude a complementary parameter space with large couplings to the light generations ($\theta$ approaching $0$). We also plot a finite naturalness `bound' in grey, for which the finite $m_h^2$ corrections exceed $\text{TeV}^2$; there remains plenty of experimentally viable parameter space that is natural, corresponding to the white region.
    }    \label{fig:W23_summary_current}
\end{figure}

In flavour, the strongest constraints come from $B$-physics. The fact that the best constrained observables probing $b \to q$ quark-flavour violating transitions, namely $B_{s(d)}$ meson mixing and $BR(B_s \to \mu\mu)$, are here all dependent on the same combination of model parameters, means we can readily infer what is the strongest bound -- which turns out to be $BR(B_s \to \mu\mu)$. We note that there are also strong constraints in the lepton sector, coming from LFUV measurements in tau decays (see Fig.~\ref{fig:W23LFV}), but these are several TeV weaker than the quark flavour constraints (assuming that the $b_L \leftrightarrow s_L$ mixing angle isn't tuned to be $\ll |V_{cb}|$ to artificially ease the bounds from $b\to s$ observables). So, for readability, we here show only the $B_s \to \mu \mu$ bound in the summary plot of Fig.~\ref{fig:W23_summary_current}. On the electroweak side, we include the bound coming from our global fit to $Z$-pole plus $m_W$ measurements (\S \ref{sec:ewFit}). For high-$p_T$ data, we similarly present the result of our combination of Drell--Yan observables measured at the LHC in all final state lepton flavour combinations (\S \ref{sec:highPT}). The interplay of these different constraints, together with the naturalness contours given in Eq. (\ref{eq:natural}), that correspond to order-1 and per-cent level tuning on $m_h^2$, are shown in Fig.~\ref{fig:W23_summary_current}. 

Taking all constraints into account, we read off the following lower bound on the mass of the $W_{23}$ triplet,
\begin{equation}
    m_{23} \gtrsim 9.4 \mathrm{~TeV}\, .
\end{equation}
This bound is saturated for values of the angular parameter around $\theta\approx\pi/4$, corresponding to $g_{1,2}=g_3$. This value of $\theta$ is not tuned or unnatural; for instance it is quite close to the value predicted in the $Sp(6)_L$ UV model discussed in \S \ref{sec:EWFU}, which predicts $\theta=\arctan (\sqrt{2})$. (We indicate this value of $\theta$ by a dotted purple line in Fig.~\ref{fig:W23_summary_current} to guide the eye).

The precise bound on $(m_{23},\theta)$ at this minimum point is driven by the EWPOs, and so LEP data still plays a decisive role in constraining this model (and similar models of this kind). That said, as one goes to slightly larger values of $g_{1,2}$, {\em e.g.} $\theta \lesssim \pi/(4.5)$, the bounds from LHC Drell--Yan observables take over. In this region of the parameter space, these high-$p_T$ LHC constraints are driven by searches in muonic channels, specifically the ATLAS single muon search (at least of the searches currently implemented in \texttt{HighPT}, that we use). Thus, currently there is great complementarity between the bounds coming from LEP electroweak precision and LHC high $p_T$ Drell--Yan observables. There is also complementarity with flavour, in particular $B_s \to \mu\mu$ branching ratios (as measured precisely by LHCb and CMS, in particular), but this bound is more `model-dependent', varying strongly with the degree of up- or down-alignment. 

Finally, let us comment more generally on the preferred parameter space of the deconstructed $SU(2)_L$ model as a whole, taking into account both the $W_{12}$ and $W_{23}$ triplets. 
We have just seen how the bound on the mass of the $W_{23}$ triplet is saturated for values of $\theta$ close to $\pi/4$, and that the $Sp(6)$ point $\theta=\arctan(\sqrt{2})$ lies close to this minimal bound.
Recall also from our discussion in \S \ref{sec:W12} that, for $\theta \approx \pi/4$, the weakest bounds on the heavier $W_{12}$ triplet are obtained for $\phi=\tan^{-1}(g_2/g_1)$ also being close to $\pi/4$. In summary, the $Sp(6)_L$-derived scenario whereby all the deconstructed gauge couplings are equal, that is $g_i=\sqrt{3} g_{SM}\, \forall i$, gives a scenario close to the lightest permissible $m_{23}$ {\em and} $m_{12}$ given current bounds, with lower bounds on the masses of order 10 TeV and 160 TeV respectively. 

Lastly, it remains to check the consistency of this parameter space point, given that $m_{ij}$ are determined as functions of the underlying vevs ($v_{ij}$) but also as functions of angles $\theta$ and $\phi$. From Eq.~\eqref{eq:Wmasses} and Fig.~\ref{fig:paramsvsangles} we see that the $Sp(6)_L$ point entails $m_{12}/m_{23} = 2/\sqrt{3} \,v_{12}/v_{23}$. To obtain masses that saturate the experimental bounds therefore implies $v_{12}/v_{23} \approx 14$, %\footnote{\textcolor{red}{*Comment 9: The Sp(6) matchpoint is precisely for when $m_{12}/m_{23} = \sqrt{2} v_{12}/v_{23}$, giving $v_{12}/v_{23} \approx 11.13$. Could possibly make reference to Joe and Joseph's EWF paper equation (4.37) where the full UV model will have additional $\mathcal{O}(1)$ corrections to our CKM.}} \textcolor{red}{*}, 
which is an $O(1)$ multiple of the ratio of CKM angles $|V_{us}|/|V_{cb}|\approx 5$. This means this parameter space point is indeed consistent with order-1 numbers appearing in the formula (\ref{eq:Vude}) for the left-handed quark mixing in our flavour model, ensuring it is a consistent model of the SM flavour structure.

\section{Prospects at Future Experiments} \label{sec:FCC}

\subsection{Improvements in Drell--Yan and flavour from High Luminosity LHC} \label{sec:HL-LHC}

By the end of the planned high-luminosity LHC (HL-LHC) phase, a target of 3 $\text{ab}^{-1}$ integrated luminosity of $pp$ collisions is expected to be accumulated. This order-of-magnitude increase in luminosity will significantly bring down the statistical uncertainty on LHC measurements. We here calculate the expected gain in sensitivity after HL-LHC in both the high $p_T$ Drell--Yan observables of \S \ref{sec:highPT}, and also for the $B_s \to \mu \mu$ branching ratio, which is the most constraining flavour observable (at least given current analyses) for our model.

For the Drell--Yan projections, we use the in-built \texttt{ChiSquareLHC} function of the \texttt{HighPT} package to rescale the $pp$ integrated luminosity up to $3  ~\text{ab}^{-1}$. This assumes the pure SM background rate is measured in all bins of the high-$p_T$ distributions, and that the statistical uncertainty on the measured event rates goes down with the square root of the gain in luminosity. As we did to obtain the current LHC bounds, we include all final state lepton channels in our computation of the projected likelihood function of the model parameters $(\theta,m_{23})$. The expected 95\% C.L. contour is then plotted in teal in Fig.~\ref{fig:FCC}.

The branching ratio of $B_s\to \mu \mu$ is the most important current flavour constraint on $W_{23}$ (see Fig.~\ref{fig:bs23}). At HL-LHC, the sensitivity is expected to improve (see~\cite{Cerri:2018ypt} Table 29). For the projections, we have used the LHCb expected precision of $4.4\%$ on the branching ratio with 400 fb$^{-1}$ integrated luminosity.\footnote{The projected exclusion could improve further if CMS achieves better than the expected precision of $7\%$ with 3 ab$^{-1}$~\cite{Cerri:2018ypt}, which seems plausible given the current precision on their branching ratio is $11\%$ with 140fb$^{-1}$~\cite{CMS:2022mgd}.} The resulting regions are shown in salmon in Fig.~\ref{fig:FCC}.

\subsection{Electroweak precision at FCC-ee} \label{sec:FCC-EW}

\begin{figure}
    \centering    
    \includegraphics[width=\textwidth]{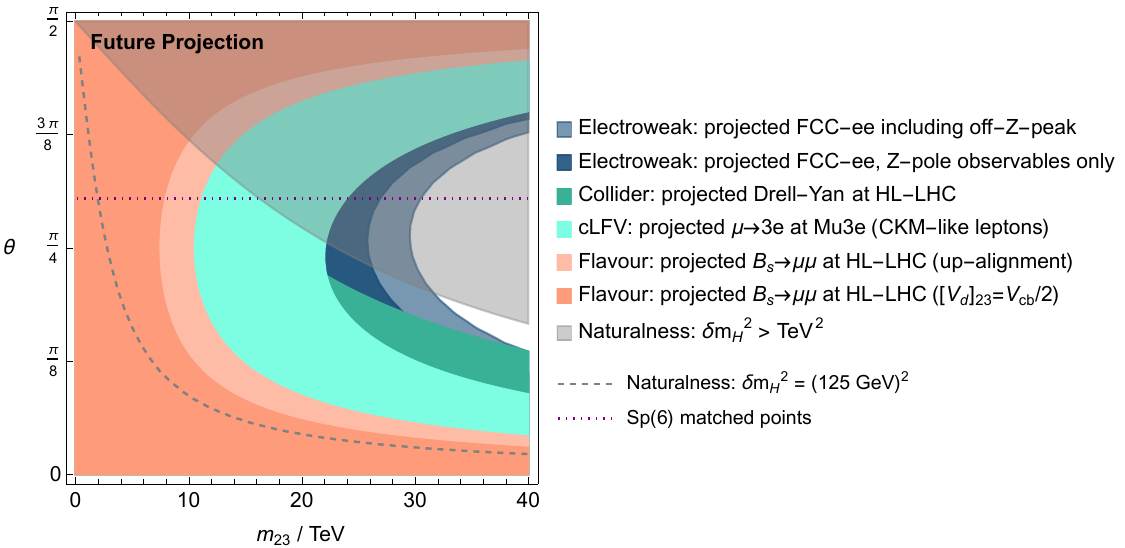}
\caption{
Projected coverage of the parameter space of the $W_{23}$ triplet, in the scenario that all measurements at various future experiments are SM-like. We include a projection of electroweak precision observables for an FCC-ee machine, accounting for the full statistics of a 4-year $Z$-pole run (\S \ref{sec:FCC-EW}), the high-luminosity LHC achieving a total integrated luminosity of 3 $\text{ab}^{-1}$ of $pp$ collisions (\S \ref{sec:HL-LHC}), and the Mu3e experiment assuming it reaches its target limit of $10^{-16}$ on the $\mu \to 3 e$ branching ratio.
}
    \label{fig:FCC}
\end{figure}

The planned $Z$ pole run at FCC-ee will be sensitive to a significant unexplored area of our parameter space, as shown by the dark blue area in Fig.~\ref{fig:FCC}. This region has been found using the projected constraints on $Z$ couplings in Table 36 of Ref.~\cite{deBlas:2022ofj}, which we have taken to be uncorrelated, as well as the projected precision on $m_W$ from Table 3 of Ref.~\cite{deBlas:2022ofj}. The paler blue region includes also the constraints on 4-fermion operators from Table 36 of the same reference, which are found from a fit including current measurements of low-energy observables as well as projected $e^+ e^- \to f \bar f$ cross-section measurements off the $Z$ pole at FCC-ee. In all cases we take the projected central value to be the SM prediction. Evidently from Fig.~\ref{fig:FCC}, the unprecedented precision of FCC-ee measurements both on and off the $Z$ pole has the power to cover much of the natural region of the general model, and could rule out that of the $Sp(6)_L$-completed model. 

FCC-ee will also be uniquely sensitive to many rare flavour processes and observables, in particular involving third generation quarks and leptons. In Appendix~\ref{sec:FCC-flavour} we consider a selection of such observables, for which dedicated sensitivity studies are available -- however, it so happens that the particular observables (in $b \to s \tau \tau$ decays, and flavour-changing top decays) receive accidentally very small BSM contributions in our model, deeming them not especially relevant despite the capabilities of FCC-ee.

\subsection{LFV and LFUV in lepton decays at Mu3e, Belle II, and FCC-ee}
\begin{figure}
    \centering
    \includegraphics[height=0.5\textwidth]{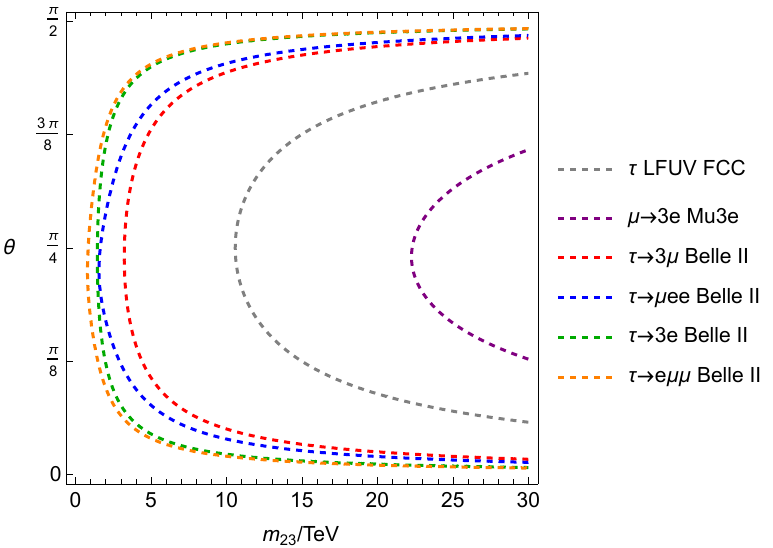}
    \caption{Projected sensitivities (90\% C.L.) of cLFV tau decays at the Belle II experiment, of LFUV tau decays at FCC-ee, and of $\mu\to 3 e$ at Mu3e, for CKM-like charged lepton mixing angles $\gamma=0.05$, $\beta=0.01$.}
    \label{fig:leptonicfuture}
\end{figure}

FCC-ee also promises to provide new insights into the nature of $\tau$ leptons, due to the enormous sample of $\tau$ pairs produced at a $Z$ pole run~\cite{Dam:2018rfz,Dam:2021ibi}. As shown in Fig.~\ref{fig:leptonicfuture}, we expect improvements in the LFUV ratios $g_\tau/g_\ell$, defined in Eqs.~\eqref{eq:tauLFUVleptonic} and \eqref{eq:tauLFUVhadronic}. Our projection has been found assuming a reduction in the total uncertainty on the measurements of a factor of 13, which is an estimate based on the total statistical plus systematic uncertainty projections on the leptonic tau branching ratios from Refs.~\cite{Dam:2018rfz,Blondel:2021ema}. The initial estimates of systematic uncertainty dominate in these projections, so if this can be reduced to the level of the statistical uncertainties, the sensitive region could expand significantly.

On this plot we also show projected future limits on cLFV $\tau$ decays from the Belle II experiment~\cite{Belle-II:2018jsg}, and on $\mu \to 3 e$ from the Mu3e experiment~\cite{Hesketh:2022wgw}. Given the expected improvement in sensitivity at Mu3e, this experiment has the best potential to observe cLFV due to the $W_{23}$ states, or to constrain the leptonic mixing angles to be smaller than CKM-like.

\subsection{Summary of Future Prospects}

The status of projected searches for the $W_{23}$ gauge bosons after FCC-ee is summarised in Fig.~\ref{fig:FCC}. Many different future experiments and measurements will probe new regions of parameter space. Most important are the great improvements anticipated in measuring $B_s\to \mu\mu$ and Drell--Yan observables at HL-LHC, and $\mu\to 3e$ at Mu3e, as well as electroweak precision observables at the FCC-ee. 

Together, these measurements will exclude nearly the entire natural parameter space of the model, which is indicated by the region outside the grey exclusion in Fig.~\ref{fig:FCC} where $|\delta m_H| > \text{TeV}$. The teal Drell--Yan regions are most sensitive where the coupling to light generations is largest, {\em i.e.} $g_1, g_2 \gg 1$, meaning regions where $\theta \to 0$. This gives these high-$p_T$ bounds a different $\theta$ dependence from the electroweak precision constraints, which also diverge when $g_3$ does since $g_3$ dictates the $W_{23}$ coupling to Higgs doublets (or equivalently the $Z_{23}-Z$ mixing). These observables therefore provide good complementarity to exclude the natural region of the model.

The quark and lepton flavour projected constraints depend on assumptions on the flavour rotation matrices of the model $V_u$ and $V_l$, which here we have taken to be CKM-like, as motivated by the explanation of the SM flavour puzzle (see \S \ref{sec:deconstruction}). But these are only determined up to order 1 factors in the absence of a UV completion, meaning the sensitivity of these flavour measurements in the $m_{23}-\theta$ plane is dependent on these factors, and so any discrepancies in these observables can impart extra experimental information about the flavour structure of the model.

\section{Conclusion}

We explore the possibility that the $SU(2)_L$ gauge interaction emerges accidentally from three separate $SU(2)_{L,i}$ forces that manifest at shorter distances, one for each family of SM fermions. This idea of deconstructing forces by flavour, which was originally conceived by Ma and collaborators in the 1980s~\cite{Li:1981nk,Li:1992fi,Ma:1987ds,Ma:1988dn,Muller:1996dj,Malkawi:1996fs} and has been recently revived in part because of anomalies in $B$-physics, offers an elegant explanation for the fermion mass hierarchies and CKM mixing pattern if the SM Higgs is charged only under $SU(2)_{L,3}$. 

The product gauge symmetry $\prod_i SU(2)_{L,i}$ will always break to its diagonal subgroup, as follows from elementary group theory arguments~\cite{Goursat1889,bauer2015generalized,Craig:2017cda}, meaning flavour universal forces emerge as almost inevitable accidents at low-energy in this setup. Going in the other direction (towards the UV), the deconstructed electroweak symmetry would follow from electroweak flavour unification deeper in the ultraviolet, whereby all three generations of left-handed fermions are unified via an $Sp(6)_L$ fundamental gauge symmetry~\cite{Davighi:2022fer}.

In this work we comprehensively analyse the phenomenology of a family-deconstructed $SU(2)_L$ gauge interaction. In doing so, we also uncover the leading phenomenological effects of the $Sp(6)_L$-unifying model. The symmetry breaking pattern yields two $SU(2)_L$ triplets of heavy gauge bosons, $W_{12}$ and $W_{23}$. The former $W_{12}$ triplet is tied to the generation of $y_1/y_2$ Yukawa hierarchies and the generation of the Cabibbo angle, and mediates flavour-violation between the light families; it must therefore be very heavy, with mass $\gtrsim 160$ TeV in order to evade light meson mixing constraints. The latter $W_{23}$ triplet is tied to the generation of $y_2/y_3$ Yukawa hierarchies and the mixing between light and third generation quarks. Its couplings are universal in the first two generations (but non-universal in the third generation), which enables it to be relatively light without contravening flavour bounds. Indeed, because the $W_{23}$ also couples directly to the SM Higgs, it gives 1-loop corrections to $m_H$; by computing these corrections and requiring they be smaller than hundreds of GeV, we delineate the parameter space of this flavour model in which the electroweak scale is natural, which roughly requires the $W_{23}$ not be heavier than tens of TeV.

This $W_{23}$ state enjoys a rich phenomenology visible across a range of experiments, with strong bounds from colliders, electroweak precision and flavour being highly complementary. In colliders, effects in Drell--Yan $pp \to \ell (\ell, \nu)$ cannot be evaded (for any lepton flavour) because each of the deconstructed gauge couplings satisfies $g_i \geq g_{SM}\approx 0.64$, so couplings to valence quarks cannot be made small despite the $U(2)$ global symmetry of the model. The electroweak fit to LEP, SLD and LHC data gives strong constraints because the $W_{23}$ talks directly to the Higgs and electroweak gauge bosons, as well as all left-handed fermions; the $W$ mass is particularly sensitive because the model predicts a negative shift in $m_W$, while experiments currently measure $m_W^{\text{exp}}>m_W^{\text{SM}}$ (even prior to the recent CDF II measurement). In flavour, we find the most constraining observable to be the $B_s \to \mu^+ \mu^-$ branching ratio, which is measured with excellent precision by the LHC experiments and for which SM theory uncertainties are under good control. 

Taking all these constraints into account, we find the bound $m_{23} \gtrsim 9.4$ TeV on the mass of the $W_{23}$ triplet (Fig.~\ref{fig:W23_summary_current}); interestingly, this lightest-allowed-mass corresponds roughly to a deconstructed $\prod_i SU(2)_{L,i}$ with equal gauge couplings $g_{1,2}=g_3$, as is predicted by the $Sp(6)_L$ UV model. We conclude with a detailed study of the prospects for probing the parameter space of the  deconstructed $SU(2)_{L,i}$ model at future experiments (Fig.~\ref{fig:FCC}). Impressive amounts of the natural parameter space will be probed by approved experiments like the High-Luminosity LHC and Mu3e. A precision EW machine like FCC-ee would, if built, explore nearly the whole natural parameter space of this model.

\section*{Acknowledgments}
%-----------------------------------------------------------------------------

We are grateful to Lukas Allwicher, Marzia Bordone, Admir Greljo, Gino Isidori, and Dave Sutherland for discussions, and to Peter Stangl for very helpful discussions about \texttt{smelli}. The work of JD was largely carried out at the University of Z\"urich, funded by the European Research Council (ERC) under the European Union’s Horizon 2020 research and innovation programme under grant agreement 833280 (FLAY), and by the Swiss National Science Foundation (SNF) under contract 200020-204428. DM and SR are partially supported by the UK Science and Technology Facilities Council (STFC) under grant ST/X000605/1. SR is supported by UKRI Stephen Hawking Fellowship EP/W005433/1. AG is supported by the UK Science and Technology Facilities Council (STFC) under grant ST/X508391/1.

\appendix

\section{Third generation quark flavour observables at FCC-ee} \label{sec:FCC-flavour}

\begin{figure}
    \centering
    \includegraphics[width=0.4\textwidth]{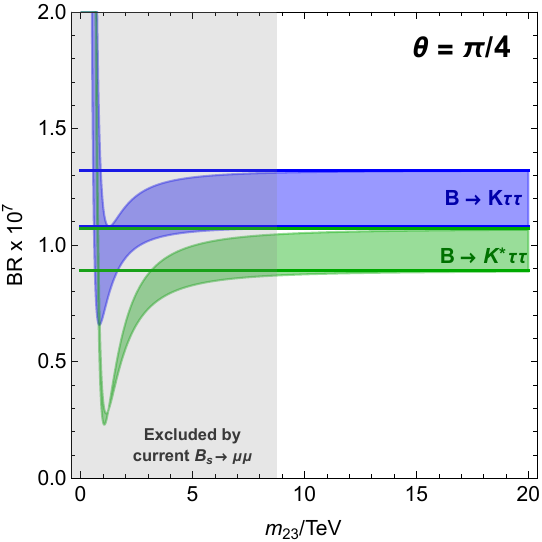}~~~~~\includegraphics[width=0.4\textwidth]{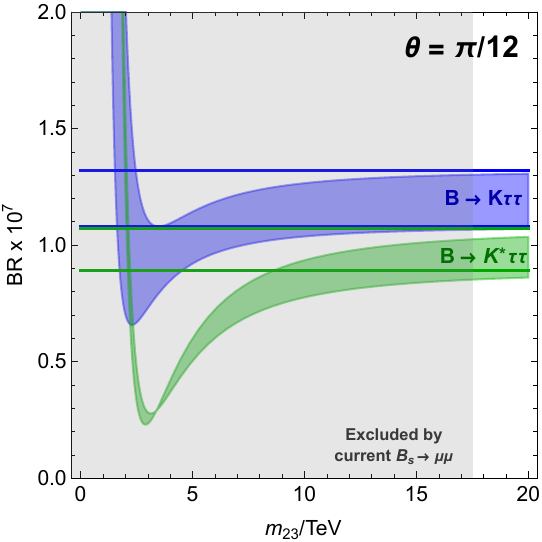}
    \caption{Branching ratios for $B\to K \tau \tau$ and $B\to K^* \tau \tau$ decays, as a function of the $W_{23}$ mass, for two different values of $\theta$. The coloured bands represent the SM+$W_{23}$ prediction within $1\sigma$ theory errors, while the horizontal lines show the $1\sigma$ boundaries of the SM-only predictions. The grey regions are excluded by current measurements of the branching ratio of $B_s\to \mu\mu$.}
    \label{fig:bstautau}
\end{figure}

\paragraph{\boldsymbol{$b\to s \tau\tau$}.}
A novelty of the FCC-ee flavour programme is the possibility of studying $b\to s \tau \tau$ decays. Many models motivated by the flavour puzzle or $B$ anomalies predict large effects here (see e.g.~\cite{Capdevila:2017iqn, Li:2020bvr,Ho:2022ipo, Allwicher:2023shc}). However, in our model the predicted change to these branching ratios are very modest, and likely unobservable even at FCC-ee. This can be seen in Fig.~\ref{fig:bstautau}, where we show the branching ratios of $B\to K^{(*)} \tau \tau$ as a function of $m_{23}$, for two different values of $\theta$ and assuming up-alignment in the quark couplings. The branching ratios and their errors have been calculated using the formulae in Ref.~\cite{Capdevila:2017iqn}, and the Wilson coefficients in Eqs.~\eqref{eq:C9tau} and \eqref{eq:C10tau}. To guide the eye, we indicate the SM-only predictions for these BRs by the horizontal lines in Fig.~\ref{fig:bstautau}. We can thus see that the parameter space regions
with large deviations from the SM predictions, which are in the low mass ($m_{23}\lesssim 3$ TeV or so) region, are already ruled out even by current $B_s\to \mu \mu $ measurements (c.f.~Fig.~\ref{fig:bs23}). We see that in the allowed region, the model in fact predicts only a tiny {\em decrease} in these $B\to K^{(\ast)}\tau\tau$ branching ratios. These decays are therefore not promising places for the $W_{23}$ to show up, especially in light of FCC-ee's projected precision on these $\tau\tau$ branching ratios: $\sim (0.5-1)\times 10^{-7}$ \cite{Li:2020bvr}. Other decay modes such as $B_s\to \phi \tau \tau$ are even less promising, and our predicted NP contribution in $B_s\to \tau \tau$ is exactly zero at tree level since we have $C_{10}^\tau=0$ (\ref{eq:C10tau}).

\begin{figure}
    \centering
    \includegraphics[width=0.7\textwidth]{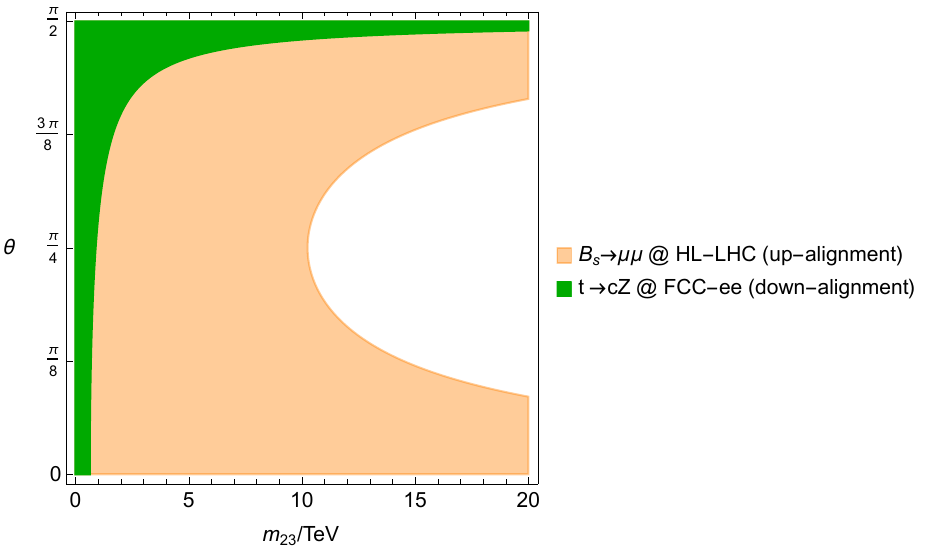}
    \caption{Projected constraints (95\% C.L.) on the parameter space of $W_{23}$ from the branching ratio of $B_s\to \mu\mu$ at LHCb at HL-LHC, and from the branching ratio of $t\to c Z$ from a $t\bar t$ run at FCC-ee.}
    \label{fig:topFCNC}
\end{figure}

\paragraph{Top FCNCs.}
If the model's quark couplings are down-aligned, then down-type FCNCs will not be present, but up-type FCNCs are unavoidable, so FCC-ee tests of e.g.~$t\to c Z$ could shed light on the model in the down-aligned limit. The sensitivity on the branching ratios of $t\to c Z$ and $t\to uZ$ decays is projected to improve to $10^{-6}$ at FCC-ee \cite{FCC:2018byv}. In our model, these decays are induced by the $O_{H q}^{(3)}$ SMEFT operator \eqref{eq:CHq3}, and the projected constraints are shown in Fig.~\ref{fig:topFCNC}, for the fully down-aligned case. Also shown is the projected constraint after HL-LHC from LHCb's measurement of the branching ratio of $B_s \to \mu \mu$~\cite{Cerri:2018ypt} in the up-aligned scenario, assuming that the measured central value remains as it is. Clearly, the top FCNCs are not projected to be very sensitive, and even if the model is fully down-aligned, the interesting region here is already ruled out by flavour-diagonal constraints such as electroweak precision.

\bibliographystyle{JHEP}
\bibliography{refs}
%%%%%%%%%%%%%%%%%%%%%%%%%%%%%%%%%%%%%%%%%%%%%%%%%%%%%%%%%
\end{document}